\documentclass[10pt,journal,compsoc]{IEEEtran}

\usepackage{times}
\usepackage{helvet}
\usepackage{courier}
\usepackage{amsfonts}
\usepackage{amsmath,bm}
\usepackage{graphicx}
\usepackage{subfigure}
\usepackage{epstopdf}
\usepackage{multirow}
\usepackage{diagbox}
\usepackage{enumerate}
\usepackage{color}
\usepackage{booktabs}

\usepackage{algorithm} 
\usepackage{algpseudocode}  
\ifCLASSOPTIONcompsoc
  \usepackage[nocompress]{cite}
\else
  \usepackage{cite}
\fi

\ifCLASSINFOpdf
\else
\fi

\begin{document}
%
\title{A Unified Framework for Community Detection and Network Representation Learning}

\author{Cunchao~Tu$^{\ast}$,
Xiangkai~Zeng$^{\ast}$,
Hao~Wang,
Zhengyan~Zhang,
Zhiyuan~Liu,
Maosong~Sun,
Bo~Zhang,
and~Leyu~Lin
\IEEEcompsocitemizethanks{\IEEEcompsocthanksitem Cunchao Tu, Hao Wang, Zhengyan Zhang, Zhiyuan Liu (corresponding author), and Maosong Sun are with the Department of Computer Science and Technology, Tsinghua University, Beijing 100084, China.\protect\\ Email: tucunchao@gmail.com, \{hwang12, zhangzhengyan14\}@mails.tsinghua.edu.cn, \{liuzy, sms\}@tsinghua.edu.cn

\IEEEcompsocthanksitem Xiangkai Zeng is with the School of Computer Science, Beihang University, Beijing 100084, China.\protect\\ Email: kevinwirehack@gmail.com


\IEEEcompsocthanksitem Bo Zhang and Leyu Lin are with the Search Product Center, WeChat Search Application Department, Tencent, Beijing 100080, China.\protect\\ Email: \{nevinzhang, goshawklin\}@tencent.com

\IEEEcompsocthanksitem $^{\ast}$ indicates equal contribution.
}
\thanks{Manuscript received April 19, 2005; revised August 26, 2015.}}

\markboth{Journal of \LaTeX\ Class Files,~Vol.~14, No.~8, August~2015}%
{Shell \MakeLowercase{\textit{et al.}}: Bare Demo of IEEEtran.cls for Computer Society Journals}

\IEEEtitleabstractindextext{%
\begin{abstract}
Network representation learning (NRL) aims to learn low-dimensional vectors for vertices in a network. Most existing NRL methods focus on learning representations from local context of vertices (such as their neighbors). Nevertheless, vertices in many complex networks also exhibit significant global patterns widely known as communities. It's intuitive that vertices in the same community tend to connect densely and share common attributes. These patterns are expected to improve NRL and benefit relevant evaluation tasks, such as link prediction and vertex classification.
Inspired by the analogy between network representation learning and text modeling, we propose a unified NRL framework by introducing community information of vertices, named as Community-enhanced Network Representation Learning (CNRL). CNRL simultaneously detects community distribution of each vertex and learns embeddings of both vertices and communities. Moreover, the proposed community enhancement mechanism can be applied to various existing NRL models.
In experiments, we evaluate our model on vertex classification, link prediction, and community detection using several real-world datasets. The results demonstrate that CNRL significantly and consistently outperforms other state-of-the-art methods while verifying our assumptions on the correlations between vertices and communities.
\end{abstract}

}

\maketitle

\IEEEdisplaynontitleabstractindextext

%
\IEEEpeerreviewmaketitle

\IEEEraisesectionheading{\section{Introduction}\label{sec:introduction}}

Network is an important way to organize relational information such as social relations, citations and other interactions among multiple objects. With the rapid development of large-scale online social networks such as Facebook, Twitter, and Weibo, network data is constantly growing and gives rise to many challenges on dealing with these large-scale real-world network data.

How to represent network data is critical when applying machine learning algorithms to network analysis tasks, such as vertex classification, personalized recommendation, anomaly detection and link prediction~\cite{shepitsen2008personalized,heard2010bayesian,liben2007link}.
Conventional graph-based representation regards each vertex as a discrete symbol, which does not consider the semantic relations between vertices and usually suffers from the sparsity issue.

In recent years, network representation learning (NRL) has been widely adopted to network analysis, which aims to build low-dimensional vectors for vertices according to their structural roles in networks. With the learnt real-valued low-dimensional representations, NRL enables us to measure the semantic relations between vertices efficiently and also alleviates the sparsity issue in conventional graph-based representation.

\begin{figure}[!htb]
\centering
\includegraphics[width=0.7\columnwidth]{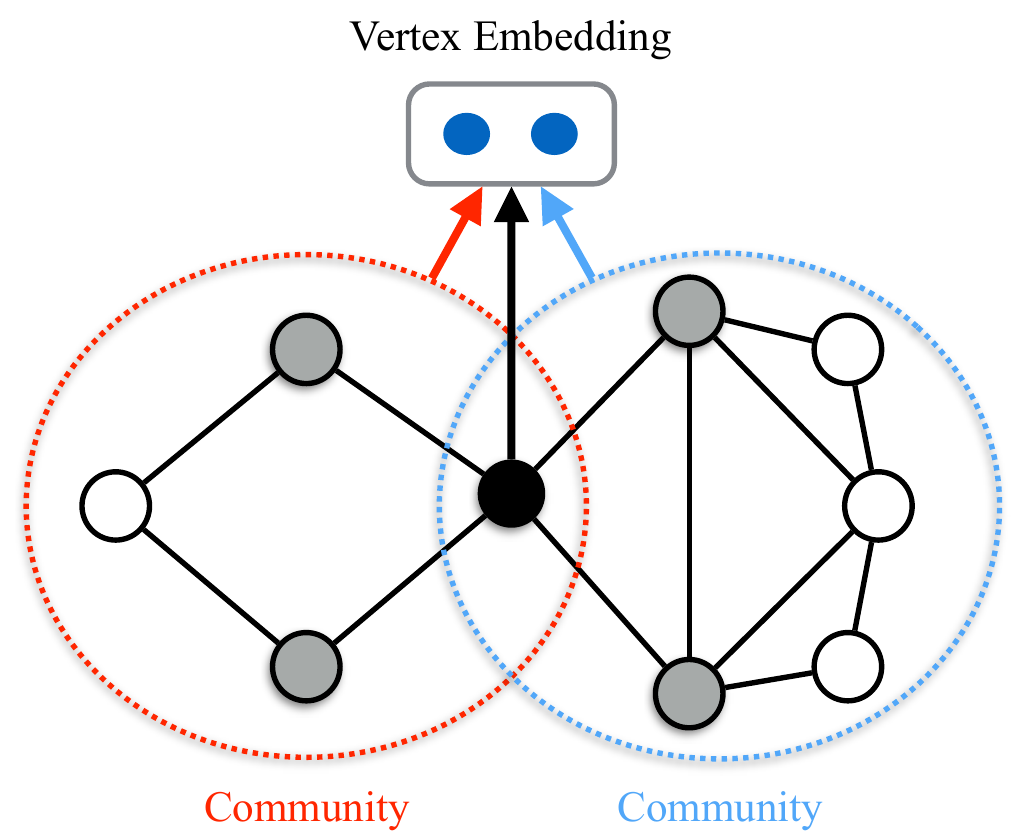}
\caption{The idea of Community-enhanced NRL: each vertex may belongs to multiple communities and the embedding of a specific vertex is determined by both its neighbor vertices and communities.}
\label{fig:idea}
\end{figure}

\begin{figure*}[!htb]
\centering
\includegraphics[width=1.5\columnwidth]{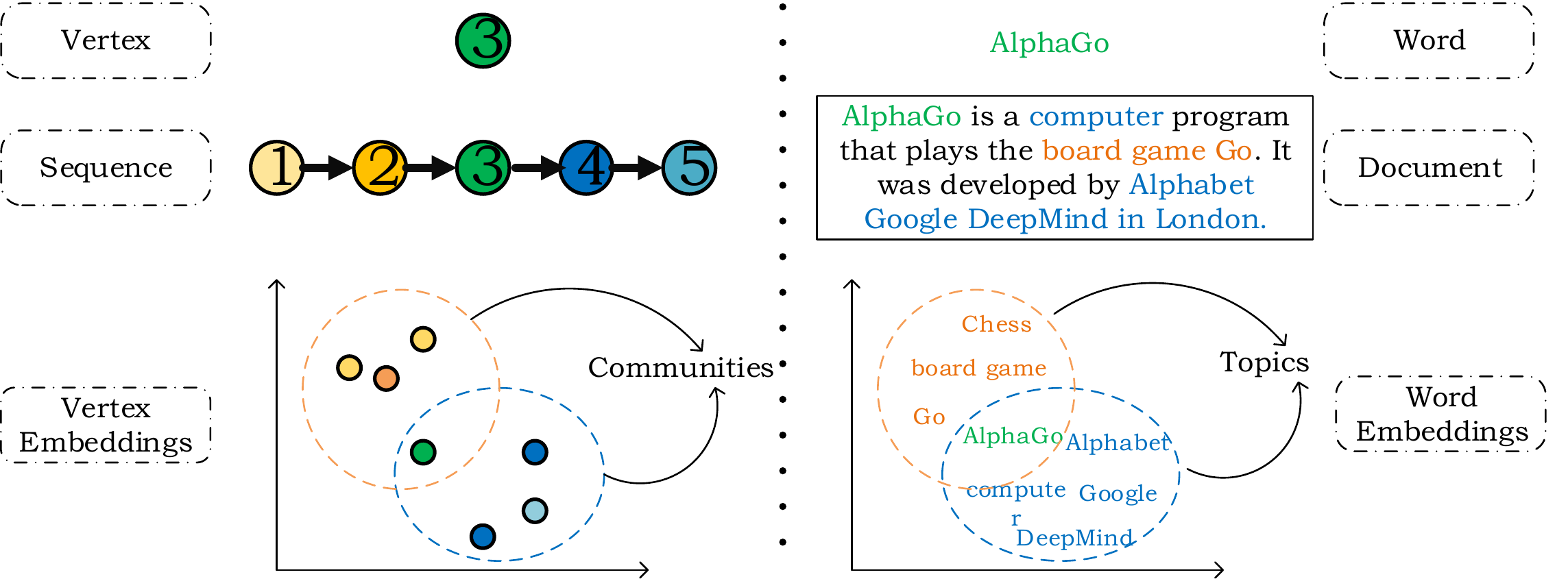}
\caption{The analogy between network representation learning and text modeling. Vertices, sequences, and communities in a network correspond to words, documents(or sentences), and topics in natural language respectively.}
\label{fig:analogy}
\end{figure*}

Most NRL methods learn vertex representations according to their \emph{local context} information. For example, DeepWalk~\cite{perozzi2014deepwalk} performs random walks over the network topology and learns vertex representations by maximizing the likelihood of predicting their local contextual vertices in walk sequences; LINE~\cite{tang2015line} models first and second order proximities of vertex pairs, and learns vertex representations by maximizing the likelihood of predicting their neighbor vertices. Both \emph{contextual vertices} in DeepWalk and \emph{neighbor vertices} in LINE are local context. Although some extensions of DeepWalk and LINE (e.g., node2vec~\cite{grover2016node2vec} and GraRep~\cite{cao2015grarep}) can explore a broader range of contextual information, the considered structural information is still limited and can only reflect local patterns.

As illustrated in Fig.~\ref{fig:idea}, in a typical complex network, vertices usually group into multiple communities with each community densely connected inside~\cite{newman2006modularity}. Vertices in a community usually share certain common attributes. For example, Facebook users with the same education-based attributes (``School name" or ``Major") tend to form communities~\cite{yang2013community}. Hence, the community structure is an important \emph{global pattern} of vertices, which is expected to benefit NRL as well as network analysis tasks. Inspired by this, we propose a unified framework for community detection and network representation learning, named as Community-enhanced NRL (CNRL).

It is worth pointing out that CNRL is highly motivated by the connection between text modeling and network modeling as well as DeepWalk~\cite{perozzi2014deepwalk}. In DeepWalk, the authors find that the vertex frequency in random walk sequences follows the power-law distribution, which is same as word frequency distribution in the text corpus. Therefore, by regarding vertices as words and vertex sequences as text, they directly re-purpose language modeling methods (i.e., skip-gram~\cite{mikolov2013distributed}) to model random walk sequences and learn vertex representations. Furthermore, as illustrated in Fig.~\ref{fig:analogy}, we extend the analogy between networks and natural languages by assuming that there exists an intermediate state between random walk sequences and vertices, i.e., communities, which corresponds to the topics between text and words. This assumption is quite intuitive and has been verified through further experiments and analysis. Based on this assumption, we propose CNRL to detect communities and learn network representations by employing topic modeling methods. Although community information has been explored by some NRL models (e.g., MNMF~\cite{wang2017community} and ComE~\cite{cavallari2017learning}), we employ an entirely different approach to solve this problem by leveraging the analogy between topics and communities. This analogy is explicit and makes CNRL easy to be integrated into random-walk based NRL models.

The basic idea of CNRL is demonstrated in Fig.~\ref{fig:framework}. We consider each vertex is grouped into multiple communities, and these communities are overlapping. Different from conventional NRL methods, where the vertex embedding is learnt from local context vertices, CNRL will learn the vertex embedding from both local context and global community information.

In CNRL, it is crucial to determine which community each vertex in a sequence belongs to. As the analogy between words in text and vertices in walk sequences has been verified by~\cite{perozzi2014deepwalk}, we assume that there are correlations between word preference on topics and vertex preference on communities as well. Following the idea of topic models, each vertex in a specific sequence is assigned with a specific community, according to the community distribution of the vertex and that of the sequence. Afterwards, each vertex and its assigned community are applied to predict its context vertices in the walk sequence. Therefore, representations of both vertices and communities are learnt by maximizing the prediction likelihood. Note that community distributions of vertices are also updated iteratively in the representation learning process.

The community representations learnt in CNRL will serve to enhance vertex representations in network analysis tasks, such as vertex classification and link prediction. More importantly, community representations are expected to benefit those long-tail vertices with less local context information.

We implement CNRL on two typical random-walk based NRL models, DeepWalk~\cite{perozzi2014deepwalk} and node2vec~\cite{grover2016node2vec}. Since both components of CNRL can be accelerated by existing parallel algorithms, it takes only linear time to train CNRL in real-world scenarios. For comparison, we conduct experiments on several real-world network datasets using the tasks of vertex classification, link prediction, and community detection.  Experimental results show that CNRL can significantly improve the performance of all the tasks, and the superiority is consistent with respect to various datasets and training ratios. It demonstrates the effectiveness of considering global community information for network representation learning.

To summarize, our major contributions are as follows:

(1) We exploit the analogy between topics in text and communities in networks, and introduce community information, a critical network analysis factor, into NRL to learn discriminative network representations. The proposed CNRL model is able to detect community structures and learn representations of vertices jointly.

(2) We propose two efficient community assignment strategies of CNRL, statistics-based one and embedding-based one. The two implementations are from different views and work equally well. Moreover, CNRL can be trained in linear due to the existing accelerated algorithms of both components.

(3) For evaluation, we conduct two typical network analysis tasks, vertex classification and link prediction on a variety of real-world networks at different scales. The experimental results show that CNRL achieves significant and consistent improvements than state-of-the-art methods.

(4) More specifically, our proposed CNRL can detect overlapping communities on multiple scales. To demonstrate the flexibility of CNRL, we visualize the detected communities of different scales and make comparisons to typical community detection method.

\section{Related Work}
Network representation learning (NRL, i.e., network embedding) aims to encode the structure and other additional information of vertices into real-valued low-dimensional representation space. Afterwards, the learnt representations will be treated as features and fed into further network analysis applications, including vertex classification, link prediction, community detection, and so on. In the following parts of this section, we will introduce related work on these highly relevant tasks and NRL respectively.

\subsection{Vertex Classification} 
As one of the most important network analysis tasks, vertex classification aims to classify each vertex into one or several categories according to its features, especially using features of network topology. It corresponds to many real-world applications, such as user profiling~\cite{Mislove2010You,li2014user}, anomaly detection~\cite{akoglu2010anomaly,hassanzadeh2012,Fire2012Strangers,Chen2012Community}, and so on.

The performance of vertex classification highly depends on the quality of vertex features. Traditional methods usually employ hand-crafted features that relevant to the category systems (e.g., age~\cite{rosenthal2011age}, gender~\cite{burger2011discriminating}, profession~\cite{tu2017prism}, personality~\cite{schwartz2013personality} et. al). However, these methods usually need many human efforts and lack of capability to other real-world scenarios.

To address these challenges, researchers propose NRL to construct effective features automatically~\cite{perozzi2014deepwalk,tang2015line}. Moreover, the low-dimensional representations in NRL models also overcome the sparsity issue and are more efficient for computation. It's also worth noting that many semi-supervised NRL models are proposed to improve the particular vertex classification task, including MMDW~\cite{tu2016max}, DDRW~\cite{li2016discriminative}, Planetoid~\cite{yang2016revisiting}, and GCN~\cite{kipf2016semi}. These methods can incorporate labeling information during the learning process, and obtain discriminative representations of vertices.

\subsection{Link Prediction} Link prediction is a fundamental task in network analysis, which aims to predict if there exists an edge between two vertices. Generally, it measures the similarity of two vertices and gives potentially connected vertices a higher score than irrelevant vertices.

Traditional methods use hand-crafted similarity measurements such as common neighbors~\cite{newman2001clustering}, Salton index~\cite{salton1986introduction}, Jaccard index and resource allocation index~\cite{zhou2009predicting} for link prediction. Besides these topology-based methods, link propagation methods~\cite{kashima2009link,raymond2010fast} are based on graph regularization, which is a common approach in semi-supervised learning. Matrix factorization methods~\cite{menon2011link} predict links by adapting matrix completion technic to the adjacency matrix. Further, people employ tensor to develop temporal model~\cite{dunlavy2011temporal} for link prediction on dynamic networks. Other methods improve the effect of link prediction by taking information like community affinity~\cite{soundarajan2012using} and vertex attribute~\cite{gong2014joint} into account.

\subsection{Community Detection}
According to the definition in~\cite{yang2010discovering}, a community in a network is a group of vertices, within which vertices are densely connected but between which they are linked sparsely. In real-world social networks, vertices in the same community usually share common properties or play similar roles~\cite{fortunato2010community}.

Detecting communities from networks is a critical research filed in social science. In terms of community detection, traditional methods focus on partitioning the vertices into different groups, i.e., detecting non-overlapping communities. Existing non-overlapping community detection works mainly include clustering-based methods~\cite{kernighan1970efficient}, modularity based methods~\cite{newman2004finding,newman2006modularity,fortunato2010community}, spectral algorithms~\cite{pothen1990partitioning}, stochastic block models~\cite{nowicki2001estimation,peixoto2015model} and so on. The major drawback of these traditional methods is that they cannot detect overlapping communities, which may not accord with real-world scenarios. To address this problem, CPM~\cite{palla2005uncovering} is proposed to generate overlapping communities by merging overlapping $k$-cliques. Ahn~\cite{ahn2010link} proposes link clustering for overlapping community detection by employing non-overlapping community detection methods to partition the links instead of vertices and then assigning a single vertex to corresponding groups of its links.

In recent years, community affiliation based algorithms show their effectiveness on overlapping community detection~\cite{wang2011community,yang2012community,xie2013overlapping,yang2013overlapping}. Community affiliation based algorithms predefine the number of communities and learn a vertex-community strength vector for each vertex and assign communities to vertices according to the vector. For example, Yang~\cite{yang2013overlapping} proposes Non-negative Matrix Factorization (NMF) method, which approximates adjacency matrix $A\in R^{|V|\times |V|}$ by $FF^T$ where matrix $F\in R^{|V|\times k}$ is vertex-community affinity matrix. Then the algorithm learns non-negative vertex embeddings and converts each dimension of the embeddings into a community. These community affiliation based algorithms try to approximate the adjacency matrix in value and design different objective functions for it. Our model follows the idea to represent the community relationship with a non-negative vector, but we don't explicitly set hard community membership for the vertices.

\subsection{Network Representation Learning}
Representation learning has shown its effectiveness in computer vision~\cite{krizhevsky2012} and natural language processing~\cite{mikolov2013distributed}. Representation learning is also becoming an important technique for network analysis in recent years. Current methods~\cite{tang2009relational,perozzi2014deepwalk,yang2015network,tang2015line} embed each vertex into a real-valued vector space based on modeling local information and take the representations as features in further evaluation tasks.

More specifically, social dimension~\cite{tang2009relational} computes the top-$d$ eigenvectors of adjacency matrix and takes them as $d$-dimensional representations. By first generating random walks from a network, DeepWalk~\cite{perozzi2014deepwalk} employs word2vec~\cite{mikolov2013distributed}, a widely-used word representation learning algorithm, to learn vertex embeddings from random walk sequences. Comparing with DeepWalk, node2vec~\cite{grover2016node2vec} designs an effective random walk strategy for sequence generation based on BFS and DFS. LINE~\cite{tang2015line} characterizes the probability of first-order and second-order proximities between two vertices. GraRep~\cite{cao2015grarep} models $k$-order proximity between two vertices through matrix factorization. Struc2vec~\cite{ribeiro2017struc2vec} constructs a multilayer context graph to encode structural similarities between vertices and generate structural context for them. Some works employ deep neural networks to learn vertex representations, such as deep auto-encoders~\cite{wangstructural,cao2016deep}, convolutional neural networks~\cite{kipf2016semi,hamilton2017inductive} and deep generation models~\cite{li2017variation}.

Vertices in real-world networks usually accompany with heterogeneous information, such as text contents and labels. There are a variety of works that attempt to incorporate this information into NRL. TADW~\cite{yang2015network} and CANE~\cite{tu2017cane} extend existing NRL models to take advantage of the accompanying text information. By utilizing labeling information of vertices, MMDW~\cite{tu2016max} employs max-margin principle to learn discriminative network representations. Chang~\cite{chang2015het} design a deep embedding architecture for capturing complex heterogeneous data in a network. TransNet~\cite{tu2017transnet} employs translation mechanism to model the interactions among vertices and labels on edges. GCN~\cite{kipf2016semi} and GraphSAGE~\cite{hamilton2017inductive} takes additional features of vertices as inputs and generate different levels of vertex representations layer-by-layer.


Most existing NRL methods only consider the local neighbors of vertices, and ignore the global patterns, such as community structure. Although random walk based methods (e.g., DeepWalk and node2vec) can explore a broad range of connected vertices and traverse in closely connected subgraphs, the exploration of contextual information is restricted by the window size, according to the proof in~\cite{yang2015network,qiu2018network}. Moreover, community information is not utilized explicitly in these models. ~\cite{wang2017community} proposes modularized nonnegative matrix factorization (MNMF) model to detect non-overlapping communities for improving vertex representations. However, there are two disadvantages of MNMF. First, it can only detect non-overlapping communities (i.e., each vertex only belongs to a specific community), which is usually not in conformity with real-world networks. Besides, MNMF is a matrix factorization based model, and its optimization complexity is $\mathcal{O}(n^2m+n^2k)$ ($n$ is the vertex number, $m$ is the dimension of vertex representation, and $k$ is the community number ), which makes it hard to handle large-scale networks.~\cite{cavallari2017learning} also proposes ComE to learn vertex embedding and detect communities simultaneously. Specifically, each community in ComE is represented as a multivariate Gaussian distribution to model how its member vertices are distributed. With well-designed iterative optimization algorithms, ComE can be trained efficiently with the complexity that is linear to the graph size. However, ComE only attempts to find the optimal multivariate Gaussian distribution of each community to fit the learnt vertex embeddings, without considering the explicit network structure of communities. In this work, we exploit the analogy between topics in text and communities in networks and propose a novel NRL model, CNRL, which can be easily and efficiently incorporated into existing NRL models. To the best of our knowledge, our model is the first attempt to learn community-enhanced network representations by utilizing the analogy between topics and communities.

\section{Community-enhanced NRL}

We start with discussing the necessary notations and formalizations of NRL.
\subsection{Formalizations}
We denote a network as $G=(V, E)$, where $V$ is the set of vertices and $E \subseteq (V\times V)$ is the set of edges, with $(v_i, v_j) \in E$ indicating there is an edge between $v_i$ and $v_j$. For each vertex $v$, NRL aims to learn a low-dimensional vector denoted as $\mathbf{v} \in \mathbb{R}^d$. Here $d$ represents the dimension of representation space.

The vertices in $G$ can be grouped into $K$ communities $C = \{c_1, \ldots, c_K\}$. The communities are usually overlapping. That is, one vertex may be the member of multiple communities in different degrees. Hence, we record the membership degree of a vertex $v$ to a community $c$ as the probability $\Pr(c|v)$, and the role of the vertex in $c$ as the probability $\Pr(v|c)$. In this work, we will also learn representations of each community $c$, denoted as $\mathbf{c}$.

In the following part, we first give a brief introduction to DeepWalk. Afterwards, we implement the idea of CNRL by extending DeepWalk to Community-enhanced DeepWalk. As the only difference between node2vec and DeepWalk is the generation methods of vertex sequences, which will not affect the extension of CNRL, we omit its implementation details.

\subsection{DeepWalk}
\begin{figure*}[t]
\centering
\includegraphics[width=1.2\columnwidth]{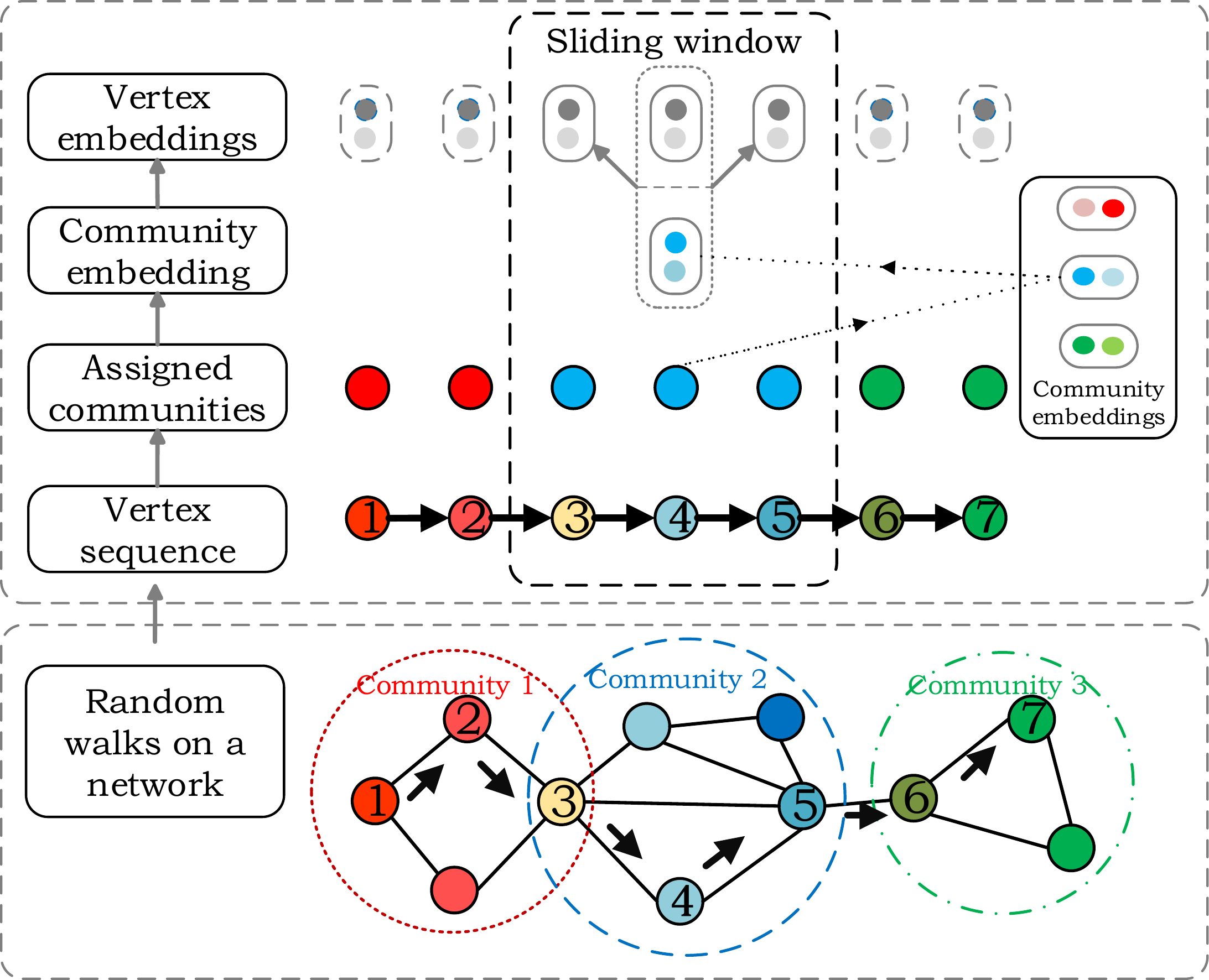}
\caption{An illustration of CNRL.}
\label{fig:framework}
\end{figure*}

DeepWalk~\cite{perozzi2014deepwalk} performs random walks over the given network $G$ firstly, and forms a set of walk sequences $S = \{s_1, \ldots, s_N\}$, where each sequence can be denoted as $s = \{v_1, \ldots, v_{|s|}\}$.

DeepWalk treats each walk sequence $s$ as a word sequence by regarding vertices as words. By introducing Skip-Gram~\cite{mikolov2013efficient}, a widely-used word representation learning algorithm, DeepWalk is able to learn vertex representations from the sequence set $S$. 

More specifically, given a vertex sequence $s = \{v_1, \ldots, v_{|s|}\}$, each vertex $v_i$ has $\{v_{i-t}, \ldots, v_{i+t}\} \setminus \{v_i\}$ as its local context vertices. Following Skip-Gram, DeepWalk learns vertex representations by maximizing the average log probability of predicting context vertices:
\begin{equation}
\mathcal{L}(s) = \frac{1}{|s|} \sum_{i=1}^{|s|} \sum_{i-t \le j \le i+t, j \ne i} \log \Pr(v_{j} | v_i),
\end{equation}
where $v_{j}$ is the context vertex of the vertex $v_i$, and the probability $\Pr(v_{j} | v_i)$ is defined by softmax function:
\begin{equation}
\Pr(v_{j} | v_i) = \frac{\exp(\mathbf{v}'_{j} \cdot \mathbf{v}_i)}{\sum_{v \in V} \exp(\mathbf{v}' \cdot \mathbf{v}_i)}.
\label{eq:w-c}
\end{equation}
Note that,  as in Skip-Gram, each vertex $v$ in DeepWalk also has two representation vectors, i.e., $\mathbf{v}_i$ when it is the center vertex and $\mathbf{v}'$ when it is the context vertex.

\subsection{Community-enhanced DeepWalk}

With random walk sequences, DeepWalk aims to maximize the local conditional probability of two vertices within a context window. That means, the co-occurrence of vertices in a sequence only relies on the affinity between vertices, while ignoring their global patterns. A critical global pattern of social networks is homophily, i.e., ``birds of a feather flock together"~\cite{mcpherson2001birds}. That is, those similar vertices sharing the same ``feather'' may group into communities. It's worth pointing out that such global pattern has not been considered by DeepWalk.

Communities provide rich contextual information of vertices. In order to take community information into consideration to provide a richer context for NRL, we make two assumptions on the correlations among vertices, walk sequences and communities.

\emph{\textbf{Assumption 1:}} Each vertex in a network may belong to multiple communities with different preferences, i.e., $\Pr(c|v)$, and each vertex sequence also owns its community distribution $\Pr(c|s)$.

With the above assumption, we make another assumption about the particular community of a vertex in a sequence.

\emph{\textbf{Assumption 2:}} A vertex in a specific walk sequence belongs to a distinct community, and the community is determined by the sequence's distribution over communities $\Pr(c|s)$ and the community's distribution over vertices $\Pr(v|c)$.

With the above assumptions and generated random walk sequences, we conduct the following two steps iteratively to detect community structures and learn representations of vertices and communities: (1) \textbf{Community Assignment.} We assign a discrete community label for each vertex in a particular walk sequence, according to both local context and global community distribution. (2) \textbf{Representation Learning.} Given a vertex and its community label, we learn optimized representations to maximize the log probability of predicting context vertices.

The two steps are demonstrated in Fig.~\ref{fig:framework}. As shown in Fig.~\ref{fig:framework}, we aim to learn an embedding for each vertex and each community. Besides, we also want to learn the community distribution of each vertex. We introduce the two steps in detail as follows.

\subsubsection{Community Assignment}
For a vertex $v$ in a walk sequence $s$, we calculate the conditional probability of a community $c$ as follows:
\begin{equation}
\Pr(c | v, s) = \frac{\Pr(c, v, s)}{\Pr(v, s)} \propto \Pr(c, v, s).
\label{eq:pc_vs}
\end{equation}
According to our assumptions, the joint distribution of $(c, v, s)$ can be formalized as
\begin{equation}
\Pr(c, v, s) = \Pr(s)\Pr(c|s)\Pr(v|c),
\label{eq:pcvs}
\end{equation}
where $\Pr(v|c)$ indicates the role of $v$ in the community $c$, and $\Pr(c|s)$ indicates the local affinity of the sequence $s$ with the community $c$. From Eq. (\ref{eq:pc_vs}) and Eq. (\ref{eq:pcvs}), we have
\begin{equation}
\Pr(c | v, s) \propto \Pr(v|c) \Pr(c|s).
\label{eq:finalp}
\end{equation}

In this work, we propose the following two strategies to implement $\Pr(c | v, s)$:

\textbf{Statistic-based assignment.} Following the Gibbs Sampling method of Latent Dirichlet Allocation (LDA)~\cite{griffiths2004finding}, we can estimate the conditional distributions of $\Pr(v|c)$ and $\Pr(c|s)$ as follows:
\begin{equation}
\Pr(v |c ) = \frac{N(v , c) + \beta}{\sum_{\tilde{v} \in V} N(\tilde{v}, c) + |V|\beta},
\label{eq:vc_count}
\end{equation}
\begin{equation}
\Pr(c |s ) = \frac{N(c , s) + \alpha}{\sum_{\tilde{c} \in C} N(\tilde{c}, s) +|K|\alpha}.
\label{eq:cs_count}
\end{equation}
Here $N(v, c)$ indicates how many times the vertex $v$ is assigned to the community $c$, and $N(c, s)$ indicates how many vertices in the sequence $s$ are assigned to the community $c$. Both $N(v, c)$ and $N(c, s)$ will be updated dynamically as community assignments change. Moreover, $\alpha$ and $\beta$ are smoothing factors following~\cite{griffiths2004finding}.

\textbf{Embedding-based assignment.} As CNRL will obtain the embeddings of vertices and communities, we can measure the conditional probabilities from an embedding view instead of global statistics. Hence, we formalize $\Pr(c|s)$ as follows:
\begin{equation}
\Pr(c|s) = \frac{\exp(\mathbf{c} \cdot \mathbf{s})}{\sum_{\tilde{c} \in C} \exp(\tilde{\mathbf{c}} \cdot \mathbf{s})},
\label{eq:cs_count2}
\end{equation}
where $\mathbf{c}$ is the community vector learnt by CNRL, and $\mathbf{s}$ is the semantic vector of the sequence $s$, which is the average of the embeddings of all vertices in $s$.
%

In fact, we can also calculate $\Pr(v|c)$ in the similar way:
 \begin{equation}
\Pr(v|c) = \frac{\exp(\mathbf{v} \cdot \mathbf{c})}{\sum_{\tilde{v} \in V} \exp(\tilde{\mathbf{v}} \cdot \mathbf{c})}.
\label{eq:vc_embedding}
\end{equation}
However, the usage of Eq. (\ref{eq:vc_embedding}) will badly degrade the performance. We suppose the reason is that vertex embedding is not exclusively learnt for measuring community membership. Hence, Eq. (\ref{eq:vc_embedding}) could not be as discriminative as compared to the statistic-based Eq. (\ref{eq:vc_count}). Therefore, in embedding-based assignment, we only calculate $\Pr(c|s)$ using embeddings and still use statistic-based $\Pr(v|c)$.

With estimated $\Pr(v|c)$ and $\Pr(c|s)$, we assign a discrete community label $c$ for each vertex $v$ in sequence $s$ according to Eq. (\ref{eq:finalp}).

\subsubsection{Representation Learning of Vertices and Communities}

Given a certain vertex sequence $s =\{v_1, \ldots, v_{|s|} \}$,  for each vertex $v_i$ and its assigned community $c_i$, we will learn representations of both vertices and communities by maximizing the log probability of predicting context vertices using both $v_i$ and $c_i$, which is formalized as follows:
\begin{equation}
\mathcal{L}(s) = \frac{1}{|s|} \sum_{i=1}^{|s|} \sum_{i-t \le j \le i+t, j \ne i} \log \Pr(v_{j} | v_i) + \log \Pr(v_{j} | c_i),
\label{eq:pred}
\end{equation}
where $\Pr(v_{j} | v_i)$ is identical to Eq. (\ref{eq:w-c}), and $\Pr(v_{j} | c_i)$ is calculated similar to $\Pr(v_{j} | v_i)$ using a softmax function:
\begin{equation}
\Pr(v_j | c_i) = \frac{\exp(\mathbf{v}'_{j} \cdot \mathbf{c}_i)}{\sum_{\tilde{v} \in V} \exp(\tilde{\mathbf{v}}' \cdot \mathbf{c}_i)}.
\label{eq:softmax}
\end{equation}

In practice, we employ negative sampling~\cite{perozzi2014deepwalk} to approximate this probability.
\subsubsection{Enhanced Vertex Representation}
After the above-mentioned representation learning process, we will obtain representations of both vertices and communities, as well as community distribution of vertex, i.e. $\Pr(c|v)$. We can apply these results to build enhanced representation for each vertex, denoted as $\hat{\mathbf v}$. The enhanced representations encode both local and global context of vertices, which are expected to promote discriminability of network representation. 

Specifically,  $\hat{\mathbf v}$ consists of two parts, the original vertex representation $\mathbf{v}$ and its community representation $\mathbf{v}_{c}$, where
\begin{equation}
\mathbf{v}_{c} = \sum_{\tilde{c} \in C} \Pr(\tilde{c}|v) \tilde{\mathbf{c}}.
\end{equation}
Afterwards, we concatenate these two parts and obtain $\hat{\mathbf v} = \mathbf{v} \oplus \mathbf{v}_{c}$.

The detailed pseudocode is shown in Algorithm~\ref{alg:cnrl}.

\makeatletter  
\def\BState{\State\hskip-\ALG@thistlm}  
\makeatother  
\begin{algorithm}  
    \begin{algorithmic}[1]  
        \Require  
        graph $G=(V, E)$, community size $K$, window size $t$
        \Ensure  
    vertex embedding $\mathbf{v}$, context embedding $\mathbf{v}'$, community distribution $\Pr(v|c)$, community embedding $\mathbf{c}$

     \State $S \gets \text{SamplePath}(G)$
    \State Initialize $\mathbf{v}$ and $\mathbf{v}'$ by Skip-Gram with $S$
        \State Assign a community for each vertex in $\mathcal{S}$ randomly
        \For {$iter = 1:L$}
        \For {each vertex $v_i$ in each sequence $s \in S$ }
        \State Calculate Eqs. (\ref{eq:vc_count}) and (\ref{eq:cs_count}) w/o current assignment
        \State Assign a community $c_i$ for  $v_i$ by Eq. (\ref{eq:finalp})
        \EndFor
        \EndFor
        \While {not convergent}
        \For {each vertex $v_i$ in each sequence $s \in S$ }
        \State Calculate Eq. (\ref{eq:finalp}) with Eq. (\ref{eq:cs_count}) or Eq. (\ref{eq:cs_count2})
        \State Assign a community $c_i$ for  $v_i$ by Eq. (\ref{eq:finalp})
        \For {each $j \in [i-t:i+t]$}
            \State SGD on $\mathbf{v}$, $\mathbf{v}'$ and $\mathbf{c}$ by Eqs. (\ref{eq:pred}) and (\ref{eq:softmax})
        \EndFor
        \EndFor
        \EndWhile
         \end{algorithmic}  
    \caption{Training Process of CNRL}
    \label{alg:cnrl}
\end{algorithm} 

\subsection{Complexity Analysis}
In CNRL, the training process consists of two parts, i.e., Gibbs Sampling of LDA and representation learning. Note that, the embedding-based assignment also requires a pre-trained LDA to calculate the statistic-based $\Pr(v|c)$. The complexity of Gibbs Sampling based LDA is $\mathcal{O}(LKnw\gamma)$, where $L$ is the loop size, $K$ is the community size, $n$ is the number of vertices, $w$ is the random walk sequences of each vertex and $\gamma$ is the sequence length. For representation learning, the complexity is $\mathcal{O}(n\log n)$ for S-DW and E-DW or $\mathcal{O}(n\log n+na^2)$ for S-n2v and E-n2v. Here, $a$ is the average degree of vertices. In total, the complexity of CNRL is $\mathcal{O}(n(LKw\gamma+\log n+a^2))$. In practice, both components of CNRL can be accelerated by parallel algorithms, such as PLDA~\cite{wang2009plda}, PLDA+~\cite{liu2011plda} and Skip-Gram~\cite{mikolov2013efficient}.

\section{Experiments}
In experiments, we adopt the tasks of vertex classification and link prediction to evaluate the performance of vertex embeddings. Besides, we also investigate the effectiveness of our model for community detection. 

\subsection{Datasets}
\begin{table}[htb]
    \centering
    \caption{Statistics of the real-world networks.}
    \label{table:datasets}
    \begin{tabular}{l|r|r|r|r}
        \toprule
        Datasets     & Cora & Citeseer & Wiki & BlogCatalog\\ \midrule
        \# Vertices & 2,708 & 3,312     & 2,405  & 10,312         \\
        \# Edges    & 5,429 & 4,732     & 15,985  & 333,983         \\
        \# Labels   & 7    & 6        & 19     & 47         \\ \midrule
        Avg.Degree & 4.01 & 2.86& 6.65    & 32.39              \\ \bottomrule
    \end{tabular}
\end{table}

We conduct experiments on four widely adopted network datasets, including Cora, Citeseer, Wiki and BlogCatalog.

\textbf{Cora.} Cora\footnote{https://people.cs.umass.edu/~mccallum/data.html} is a research paper set constructed by \cite{McCallumIRJ}. It contains $2,708$ machine learning papers which are categorized into $7$ classes. The citation relationships between them form a typical social network.

\textbf{Citeseer.} Citeseer is another research paper set constructed by \cite{McCallumIRJ}. It contains $3,312$ publications and $4,732$ connections between them. These papers are from $6$ classes.

{\textbf{Wiki.}} Wiki \cite{sen2008collective} contains $2,405$ web pages from $19$ categories and $15,985$ links between them. It's much denser than Cora and Citeseer.

{\textbf{BlogCatalog.}} BlogCatalog~\cite{tang2009relational} is a social network among blog authors. The bloggers also provide their interested topics, which are regarded as labels.

More detailed information about these datasets is listed in Table~\ref{table:datasets}. These networks belong to different categories and own certain characteristics, such as sparsity and amount of labels.

Besides, we also conduct experiments on a toy network named Zachary's Karate network~\cite{zachary1977information} to visualize detected communities by our model. Karate network is a social network of friendships between a karate club members. It contains $34$ vertices and $78$ edges.

\begin{figure}[!htb]
\centering
\includegraphics[width=0.8\columnwidth]{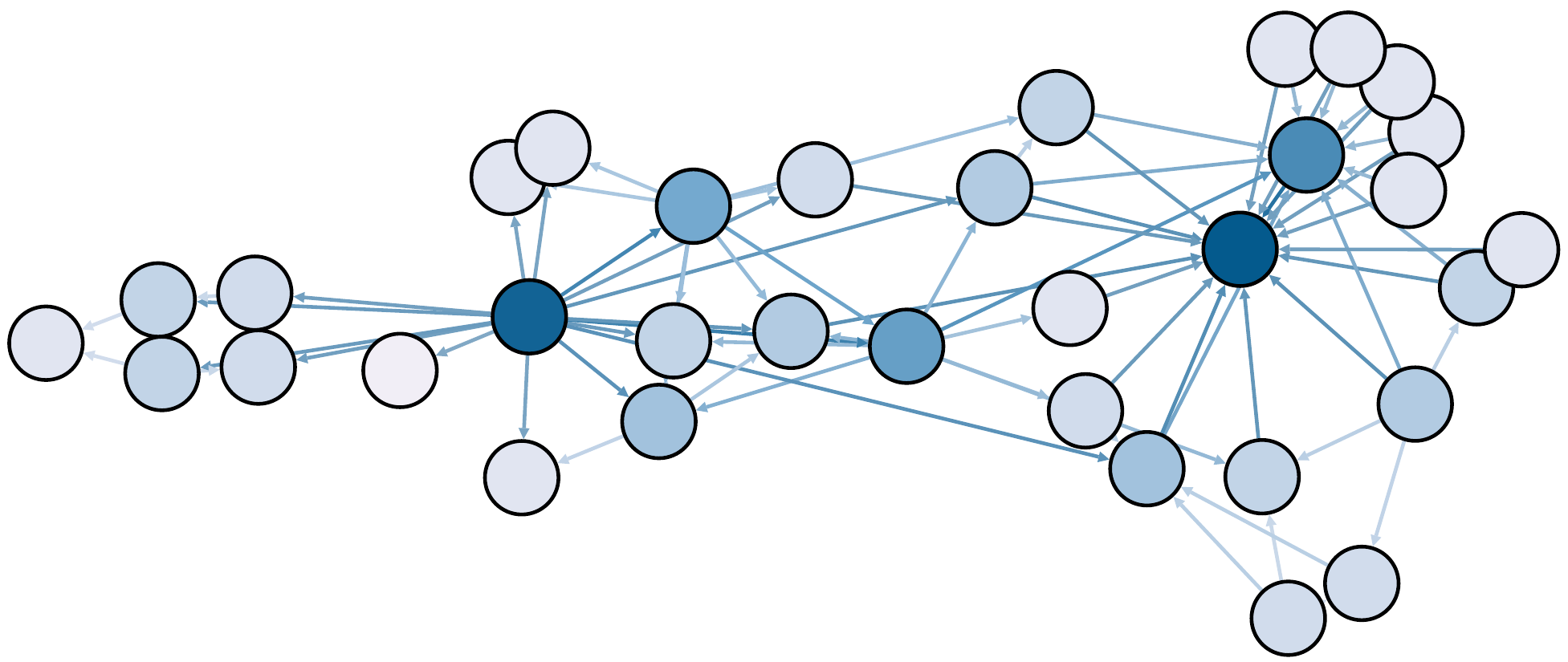}
\caption{Karate network. The color of each vertex indicates the its degree. The stronger the background color is, the larger the degree of this vertex is.}
\label{fig:karate}
\end{figure}

\subsection{Baseline Methods}

We employ six state-of-the-art NRL models as baselines, including \textbf{DeepWalk}, \textbf{LINE}, \textbf{node2vec}, \textbf{SDNE}, \textbf{MNMF}, and \textbf{ComE}.  

{\textbf{DeepWalk:}} DeepWalk is an efficient network representation learning model proposed by \cite{perozzi2014deepwalk}. It performs random walks on networks to generate vertex sequences. With these sequences, DeepWalk adopts a widely used word representation learning model, Skip Gram \cite{mikolov2013efficient}, to learn vertex representations.

{\textbf{LINE:}} LINE~\cite{tang2015line} is another network representation learning model that can handle large-scale social networks with millions of vertices. In LINE, the first-order proximity and second-order proximity are proposed to represent the joint and conditional probabilities of vertex pairs respectively. In experiments, we employ both first-order (denoted as LINE-1st) and second-order (denoted as LINE-2nd) LINE as baselines.

{\textbf{node2vec:}} Node2vec~\cite{grover2016node2vec} is an extension of DeepWalk. With the utilization of BFS and DFS strategies, node2vec designs a biased random walk strategy to generate vertex sequences and achieves significant improvements than DeepWalk.

{\textbf{SDNE:}} SDNE~\cite{wangstructural} is the first attempt to employ deep neural networks (i.e., deep auto-encoder) to construct vertex embeddings. 

{\textbf{MNMF:}} MNMF~\cite{wang2017community} jointly detects non-overlapping communities and learns network representations through matrix factorization. The learnt representations are expected to preserve the community information.

{\textbf{ComE:}} ComE~\cite{cavallari2017learning} learns network representation and detects overlapping  communities jointly. Specifically, it employs a multivariate Gaussian distribution to represent each community.

Besides, we also employ four popular link prediction methods as baselines, which are mainly based on local topological properties~\cite{lu2011link}:

{\textbf{Common Neighbors (CN).}} For vertex $v_i$ and $v_j$, CN~\cite{newman2001clustering} simply counts the common neighbors of $v_i$ and $v_j$ as similarity score:
\begin{equation}
sim(v]_i,v_j)=|N_i^+ \cap N_j^+|.
\end{equation}

{\textbf{Salton Index.}} For vertex $v_i$ and $v_j$, Salton index~\cite{salton1986introduction} further considers the degree of $v_i$ and $v_j$. The similarity score can be formulated as:
\begin{equation}
sim(v_i,v_j)=(|N_i^+ \cap N_j^+|)/(\sqrt{|N_i^+|\times |N_j^+|}).
\end{equation}

{\textbf{Jaccard Index.}} For vertex $v_i$ and $v_j$, Jaccard index is defined as:
\begin{equation}
sim(v_i,v_j)=(|N_i^+ \cap N_j^+|)/(|N_i^+ \cup N_j^+|).
\end{equation}

{\textbf{Resource Allocation Index (RA).}} RA index~\cite{zhou2009predicting} is the sum of resources received by $v_j$:
\begin{equation}
sim(v_i,v_j)=\sum_{v_k\in N_i^+}\frac{1}{|N_k^+|}.
\end{equation}

For community detection, we select four most prominent methods as baselines:

{\textbf{Sequential Clique Percolation (SCP)}} ~\cite{kumpula2008sequential} is a faster version of Clique Percolation~\cite{palla2005uncovering}, which detects communities by searching for adjacent cliques.

{\textbf{Link Clustering (LC)}}~\cite{ahn2010link} aims to find link communities rather than nodes.

{\textbf{MDL}}~\cite{peixoto2015model} employs the minimum description length principle (MDL) to perform model selection over various stochastic block models.

{\textbf{BigCLAM}}~\cite{yang2012community} is a typical nonnegative matrix factorization based model which can detect densely overlapping and hierarchically nested communities in massive networks.

\begin{table*}[!htb]
\centering
\caption{Vertex classification accuracies (\%) on Cora}
\label{table:accuracy_cora}
\begin{tabular}{l|l|l|l|l|l|l|l|l|l|l}
    \toprule
    \multicolumn{2}{c|}{\%Training ratio}     & 10\%           & 20\%           & 30\%           & 40\%           & 50\%           & 60\%           & 70\%           & 80\%           & 90\%           \\\midrule
    \multicolumn{2}{c|}{DeepWalk}      & $70.77$ & $73.35$ & $74.53$ & $74.94$ & $75.62$ & $76.07$ & $76.08$ & $76.33$ & $77.27$\\
    \multicolumn{2}{c|}{LINE}      & $70.61$ & $74.79$ & $76.93$ & $77.99$ & $78.66$ & $79.22$ & $79.53$ & $79.35$ & $79.67$\\
    \multicolumn{2}{c|}{node2vec ($p=2, q=0.5$)}      & $73.29$ & $76.03$ & $ 77.52 $ & $ 78.08 $ & $ 78.40 $ & $ 78.51 $ & $ 78.72 $ & $ 79.06 $ & $ 79.15 $\\
        \multicolumn{2}{c|}{SDNE}      & $70.97$ & $75.08$ & $ 76.90 $ & $ 77.82 $ & $ 78.26 $ & $ 79.11 $ & $  79.37  $ & $ 79.46 $ & $ 79.37 $\\ 
    \multicolumn{2}{c|}{MNMF}      & $ 75.08 $ & $ 77.85 $ & $ 79.05 $ & $ 79.53 $ & $ 79.82 $ & $ 80.21 $ & $ 79.98 $ & $ 80.11 $ & $ 79.41 $\\
        \multicolumn{2}{c|}{ComE}      & $\mathbf{ 76.72}$ & $  79.25  $ & $  80.73  $ & $ 80.97 $ & $  81.53  $ & $  82.10  $ & $  82.19  $ & $ 82.42 $ & $82.65$\\
    \midrule
    \multicolumn{2}{c|}{\textbf{S-DW}} & $72.78$ & $75.93$ & $77.47$ & $77.98$ & $78.69$ & $79.14$ & $79.15$ & $78.99$ & $78.23$\\
    \multicolumn{2}{c|}{\textbf{E-DW}} & $73.35$ & $76.56$ & $77.11$ & $78.63$ & $79.18$ & $79.86$ & $79.96$ & $79.94$ & $80.26$\\ \midrule
    \multicolumn{2}{c|}{\textbf{S-n2v} ($p=4, q=1$)} & $ 75.86 $ & $ \mathbf{79.92} $ & $\mathbf{ 81.21 }$ & $ \mathbf{ 82.13} $ & $\mathbf{  82.81} $ & $\mathbf{  83.06} $ & $ \mathbf{ 82.95} $ & $ \mathbf{ 83.78} $ & $ \mathbf{ 83.65} $\\
    \multicolumn{2}{c|}{\textbf{E-n2v} ($p=4, q=1$)} & $76.30$ & $79.40$ & $ 80.62 $ & $81.19 $ & $ 81.46 $ & $ 81.82 $ & $ 81.67 $ & $ 82.16$ & $ 82.80 $\\ \bottomrule
\end{tabular}
\end{table*}
\begin{table*}[!htb]
    \centering
    \caption{Vertex classification accuracies (\%) on Citeseer}
    \label{table:accuracy_citeseer}
    \begin{tabular}{l|l|l|l|l|l|l|l|l|l|l}
        \toprule
        \multicolumn{2}{c|}{\%Training ratio}     & 10\%           & 20\%           & 30\%           & 40\%           & 50\%           & 60\%           & 70\%           & 80\%           & 90\%           \\ \midrule
        \multicolumn{2}{c|}{DeepWalk}      & $47.92$ & $ 51.54 $ & $52.92$ & $54.14$ & $54.21$ & $54.58$ & $55.07$ & $56.09$ & $55.33$\\
        \multicolumn{2}{c|}{LINE}      & $44.27$ & $47.57$ & $50.10$ & $51.15$ & $51.93$ & $52.74$ & $53.46$ & $53.98$ & $54.01$\\
        \multicolumn{2}{c|}{node2vec ($p=2, q=0.25$)}      & $ 49.47 $ & $ 53.27 $ & $ 54.22 $ & $ 55.51 $ & $ 55.87 $ & $ 56.34 $ & $ 56.95 $ & $ 57.61 $ & $ 57.56 $\\
        \multicolumn{2}{c|}{SDNE}      & $  47.35  $ & $  51.10 $ & $  52.45  $ & $  53.20  $ & $  53.70  $ & $  54.20  $ & $  54.79  $ & $  55.26  $ & $  54.46  $\\
        \multicolumn{2}{c|}{MNMF}      & $ 51.62 $ & $ 53.80 $ & $ 55.47 $ & $ 56.94 $ & $ 56.81 $ & $ 57.04 $ & $ 57.05 $ & $ 57.00 $ & $ 57.22 $\\ 
        \multicolumn{2}{c|}{ComE}      & $  \mathbf{54.71}  $ & $  \mathbf{57.70}  $ & $\mathbf{58.84}  $ & $\mathbf{59.67}  $ & $  59.93$ & $ 60.30$ & $  61.12 $ & $  61.62$ & $  61.11$\\\midrule
        \multicolumn{2}{c|}{\textbf{S-DW}} & $49.40$ & $52.58$ & ${54.83}$ & ${55.92}$ & ${56.63}$ & ${56.99}$ & ${57.46}$ & ${58.48}$ & ${58.14}$\\
\multicolumn{2}{c|}{\textbf{E-DW}} & ${49.48}$ & $52.52$ & $54.41$ & $55.29$ & $56.25$ & $56.21$ & $57.14$ & $57.53$ & $57.41$\\ \midrule
    \multicolumn{2}{c|}{\textbf{S-n2v} ($p=4, q=1$)} & $  53.12  $ & $ 56.68  $ & $58.20$ & $ 59.48$ & $  \mathbf{60.31}  $ & $ \mathbf{ 60.76 } $ & $ \mathbf{ 61.21}  $ & $  \mathbf{61.90}  $ & $  \mathbf{62.63}  $\\
    \multicolumn{2}{c|}{\textbf{E-n2v} ($p=4, q=1$)} & ${  51.84  }$ & ${  54.33  }$ & $  55.85  $ & ${  56.47  }$ & ${  57.19  }$ & ${  57.11  }$ & ${  57.85  }$ & ${  58.67  }$ & ${  58.51  }$\\ \bottomrule
 \end{tabular}
\end{table*}
\begin{table*}[!htb]
    \centering
    \caption{Vertex classification accuracies (\%) on Wiki}
    \label{table:accuracy_wiki}
    \begin{tabular}{l|l|l|l|l|l|l|l|l|l|l}
        \toprule
        \multicolumn{2}{c|}{\%Training ratio}     & 10\%           & 20\%           & 30\%           & 40\%           & 50\%           & 60\%           & 70\%           & 80\%           & 90\%           \\ \midrule
        \multicolumn{2}{c|}{DeepWalk}      & $ 58.54 $ & $ 62.12 $ & $ 63.56 $ & $65.22$ & $65.90$ & $66.53$ & $67.22$ & $67.50$ & $67.56$\\
        \multicolumn{2}{c|}{LINE}      & $57.53$ & $61.47$ & $63.45$ & $65.14$ & $66.55$ & $67.66$ & $68.35$ & $68.21$ & $68.31$\\
        \multicolumn{2}{c|}{node2vec ($p=1, q=0.5$)}      & $ 58.93 $ & $ 62.60 $ & $ 64.11 $ & $ 65.36 $ & $ 66.03 $ & $ 67.38 $ & $ 67.93 $ & $ 68.26 $ & $ 68.99 $\\
        \multicolumn{2}{c|}{SDNE}     & $ 52.42 $ & $ 57.34 $ & $ 60.15 $ & $62.35 $ & $63.18 $ & $64.21 $ & $ 64.71 $ & $65.63 $ & $65.60$  \\
        \multicolumn{2}{c|}{MNMF}      & $ 54.76 $ & $ 58.82 $ & $ 60.43 $ & $ 61.66 $ & $ 62.74 $ & $ 63.23 $ & $63.46$ & $ 63.45 $ & $ 64.77 $\\ 
        \multicolumn{2}{c|}{ComE}      & $ 59.11 $ & $62.46 $ & $64.38 $ & $65.45 $ & $65.98 $ & $67.38 $ & $67.49 $ & $67.92 $ & $67.89 $ \\\midrule
        \multicolumn{2}{c|}{\textbf{S-DW}} & ${59.97}$ & ${63.41}$ & ${65.48}$ & ${67.03}$ & ${67.95}$ & ${69.15}$ & ${69.45}$ & ${70.03}$ & ${70.63}$\\
\multicolumn{2}{c|}{\textbf{E-DW}} & $58.69$ & $62.37$ & $64.01$ & $65.46$ & $66.16$ & $66.85$ & $66.79$ & $67.05$ & $67.05$\\ \midrule
    \multicolumn{2}{c|}{\textbf{S-n2v} ($p=2, q=1$)} & $  \mathbf{ 60.66}  $ & $  \mathbf{ 64.43}  $ & $\mathbf{  66.63  }$ & $  \mathbf{ 68.23}  $ & $  \mathbf{ 68.92}  $ & $  \mathbf{ 70.52}  $ & $  \mathbf{ 70.41}  $ & $  \mathbf{ 70.62 } $ & $  \mathbf{ 71.60}  $\\
    \multicolumn{2}{c|}{\textbf{E-n2v} ($p=1, q=0.5$)} & ${   60.07   }$ & ${  63.81  }$ & $  65.52  $ & ${  66.69  }$ & ${  67.64  }$ & ${  69.21  }$ & ${  69.62  }$ & ${  69.40  }$ & ${  70.71  }$\\ \bottomrule
    \end{tabular}
\end{table*}

\begin{table*}[!htb]
    \centering
    \caption{Vertex classification results (\%) on BlogCatalog (micro-precision).}
    \label{table:accuracy_blog}
    \begin{tabular}{l|l|l|l|l|l|l|l|l|l|l}
        \toprule
        \multicolumn{2}{c|}{\%Training ratio}     & 1\%           & 2\%           & 3\%           & 4\%           & 5\%           & 6\%           & 7\%           & 8\%           & 9\%           \\ \midrule
        \multicolumn{2}{c|}{DeepWalk}      & $ 23.66 $ & $ 27.12 $ & $ 28.28 $ & $30.02$ & $30.58$ & $31.37$ & $31.57$ & $31.71$ & $32.31$\\
        \multicolumn{2}{c|}{LINE}      & $19.31$ & $23.21$ & $22.88$ & $24.82$ & $25.89$ & $27.00$ & $27.75$ & $28.70$ & $30.04$\\
        \multicolumn{2}{c|}{node2vec}      & $ 24.47 $ & $ 27.83 $ & $ 29.11 $ & $ 30.61 $ & $ 30.87 $ & $ 31.05 $ & $ 31.50 $ & $ 31.44 $ & $ 31.96 $\\
        \multicolumn{2}{c|}{SDNE}     & $ 17.73 $ & $ 22.38 $ & $ 23.92 $ & $ 25.06 $ & $ 25.65 $ & $ 27.05 $ & $ 27.44 $ & $ 27.72 $ & $ 27.97 $ \\
        \multicolumn{2}{c|}{MNMF}      & $ 19.26 $ & $ 21.08 $ & $ 22.29 $ & $ 23.99 $ & $25.24 $ & $ 25.97 $ & $26.31$ & $ 26.58 $ & $ 27.16 $\\
        \multicolumn{2}{c|}{ComE}     & $ 22.67 $ & $ 27.43 $ & $ 28.49 $ & $ 29.79 $ & $ 30.34 $ & $ 31.04 $ & $ 31.32 $ & $ 31.58 $ & $ 32.15 $\\ \midrule
        \multicolumn{2}{c|}{\textbf{S-DW}} & ${23.80}$ & ${27.02}$ & ${28.63}$ & ${30.14}$ & ${30.25}$ & ${30.96}$ & ${31.16}$ & ${31.46}$ & ${32.45}$\\
\multicolumn{2}{c|}{\textbf{E-DW}} & $24.93$ & $\mathbf{28.36}$ & $29.28$ & $\mathbf{30.80}$ & $\mathbf{31.19}$ & $31.65$ & $31.72$ & $\mathbf{32.22}$ & $\mathbf{32.76}$\\ \midrule
    \multicolumn{2}{c|}{\textbf{S-n2v}} & $  24.95  $ & $  27.88  $ & ${29.17}$ & $  30.24  $ & $  30.95  $ & $  \mathbf{31.77}  $ & $  \mathbf{31.85}  $ & $  32.12  $ & $  32.34  $\\
    \multicolumn{2}{c|}{\textbf{E-n2v}} & $\mathbf{25.75}$ & ${28.29}$ & $  \mathbf{29.36}  $ & ${ 30.54  }$ & ${ 31.13  }$ & ${  31.36  }$ & ${  31.67  }$ & ${  31.99  }$ & ${  32.60  }$\\ \bottomrule
    \end{tabular}
\end{table*}
\begin{figure*}[!htb]
\centering
\begin{minipage}{\textwidth}
\subfigure[Classification accuracies on Cora]{
\includegraphics[width=0.31\textwidth,height=1.8in]{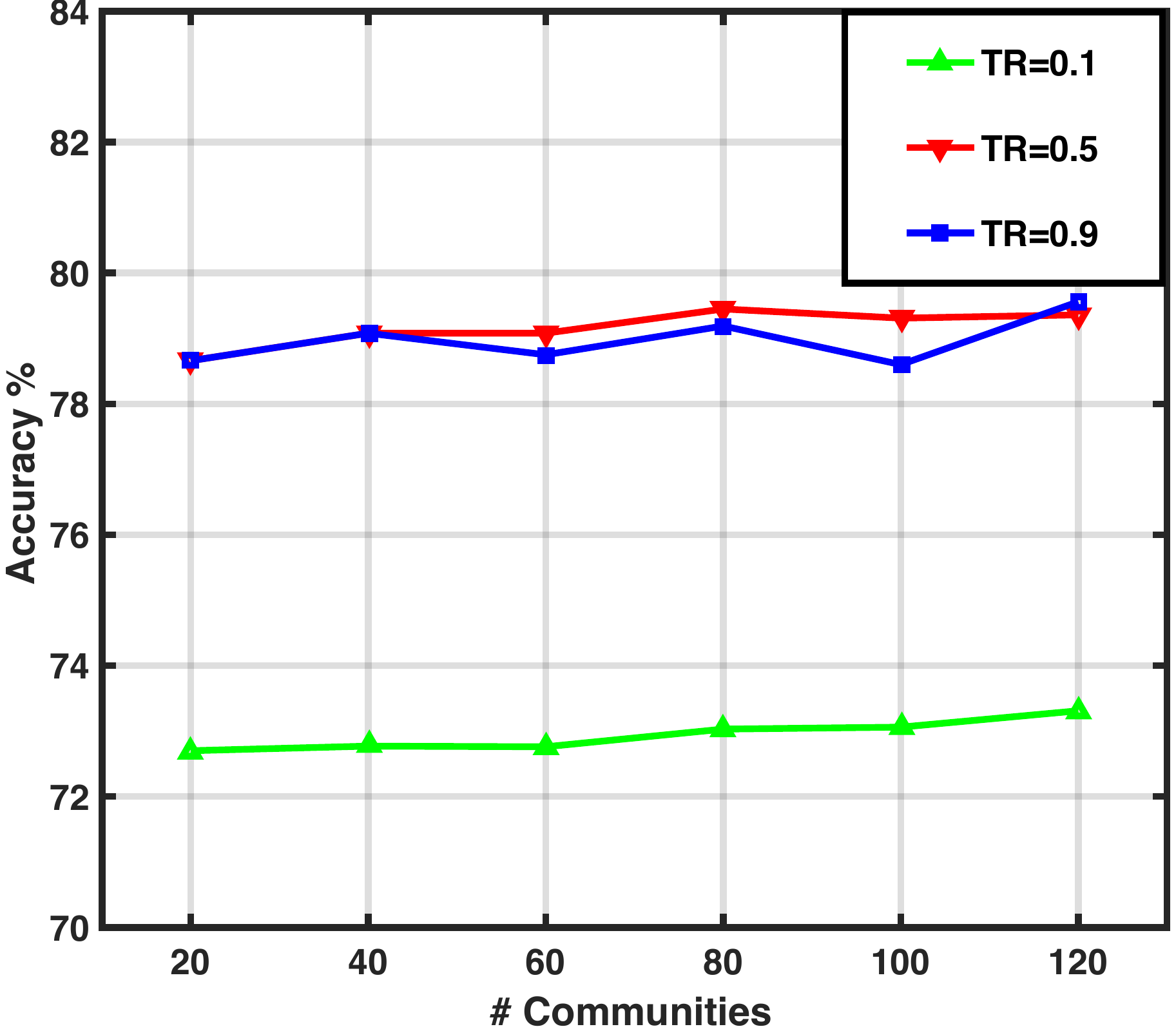}
}
\subfigure[Classification accuracies on Citeseer]{
\includegraphics[width=0.31\textwidth,height=1.8in]{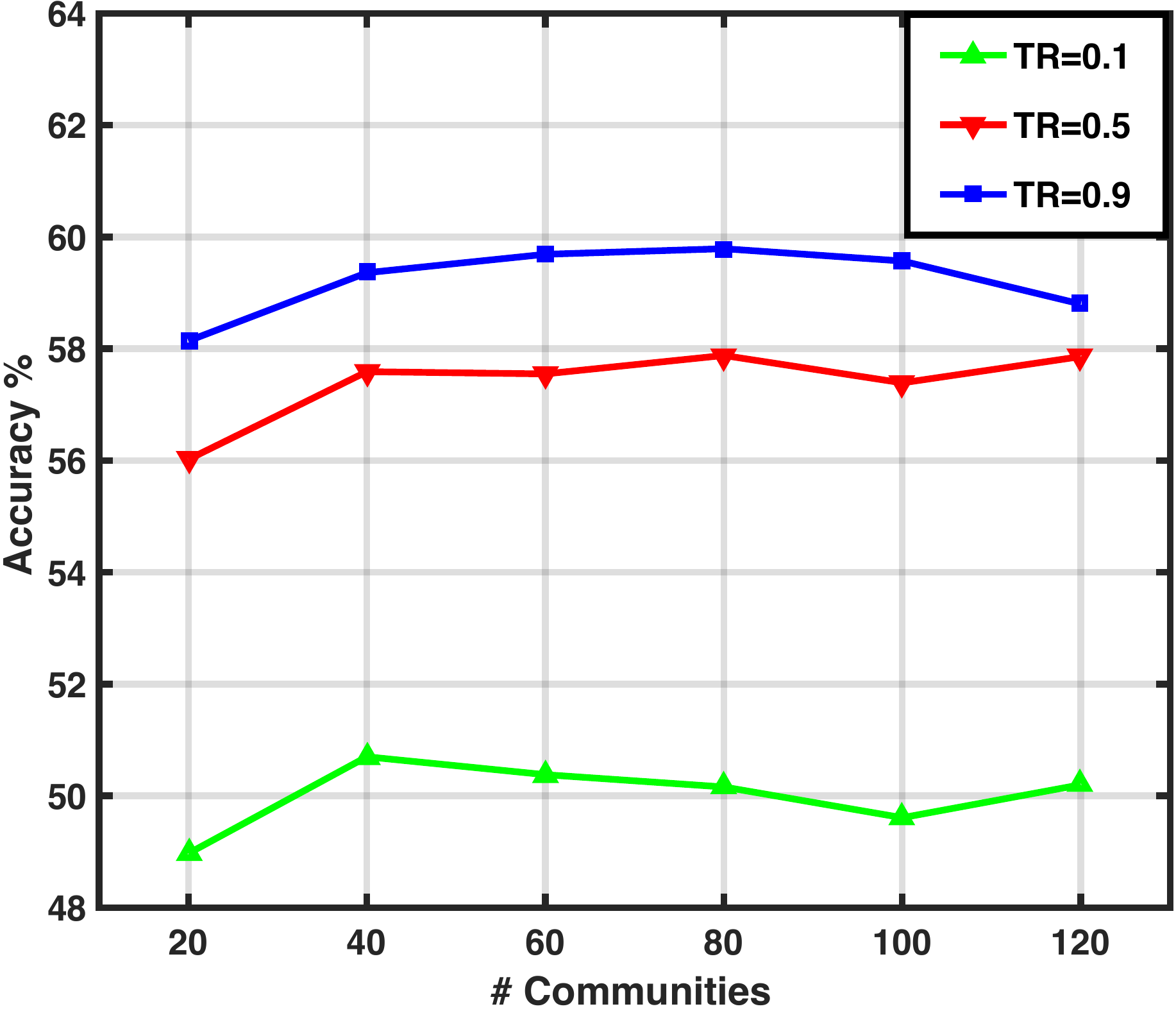}
}
\subfigure[Classification accuracies on Wiki]{
\includegraphics[width=0.31\textwidth,height=1.8in]{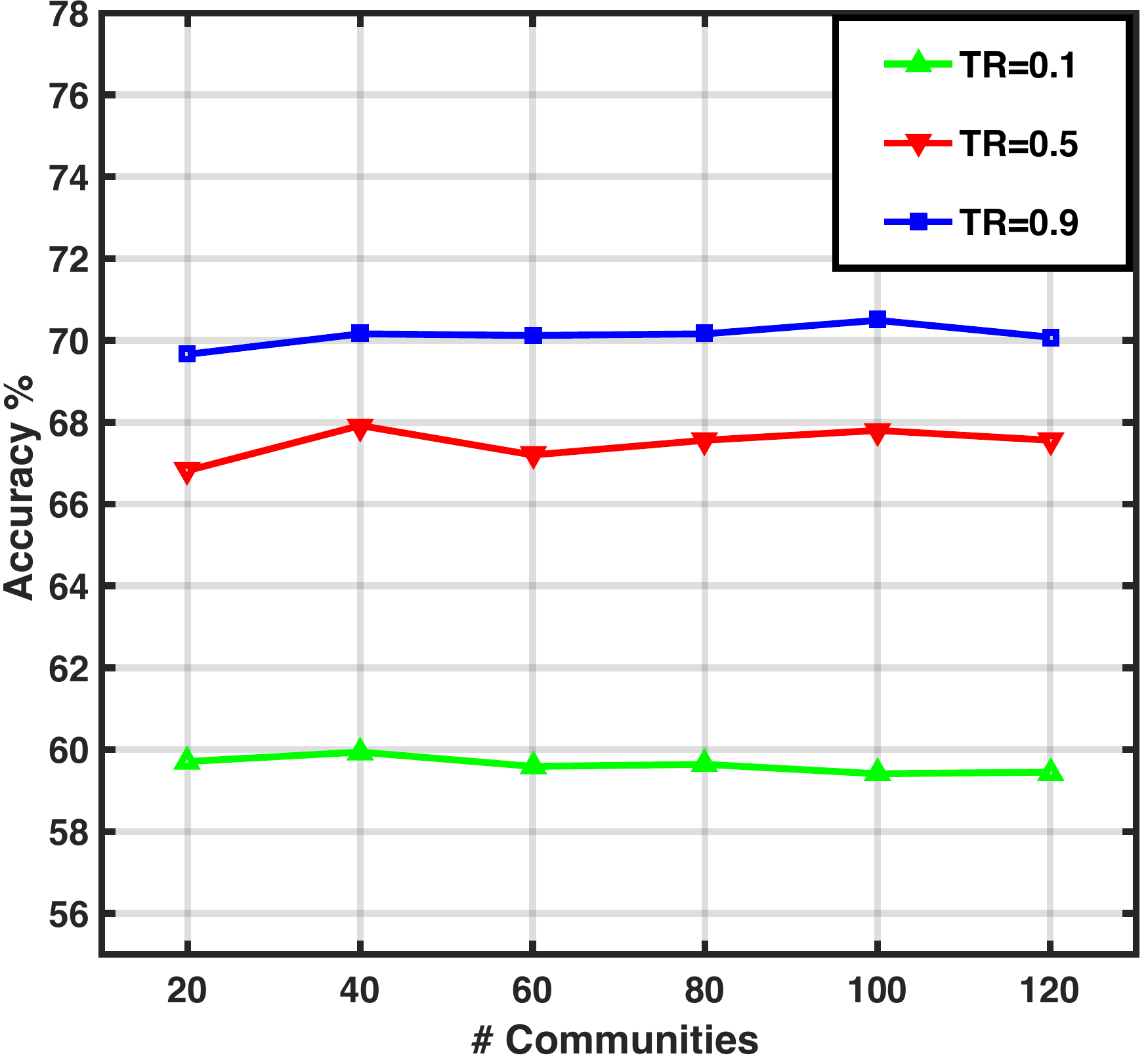}
}
\end{minipage}
\caption{Parameter sensitivity on vertex classification. (TR denotes training ratio.)}
\label{fig:vc_sen}
\end{figure*}

\subsection{Parameter Settings and Evaluation Metrics}
As mentioned in the previous section, we implement CNRL on DeepWalk and node2vec. Taking DeepWalk for example, we denote the statistic-based and embedding-based community-enhanced DeepWalk as S-DW and E-DW respectively. Similarly, the corresponding implementations of node2vec are denoted as S-n2v and E-n2v.

For a fair comparison, we apply the same representation dimension as $128$ in all methods. In LINE, as suggested in~\cite{tang2015line}, we set the number of negative samples to $5$ and the learning rate to $0.025$. We set the number of side samples to $1$ billion for BlogCatalog and $10$ million for other datasets. For random walk sequences, we set the walk length to $40$, the window size to $5$ and the number of sequences started from each vertex to $10$. Besides, we employ grid-search to obtain the best performing hyper-parameters in MNMF.

Note that, the representation vectors in CNRL consist of two parts, including the original vertex vectors and the corresponding community vectors. For a fair comparison, we set the dimension of both vectors to $64$ and finally obtain a 128-dimensional vector for each vertex. Besides, the smoothing factor $\alpha$ is set to $2$ and $\beta$ is set to $0.5$.

For vertex classification, as each vertex in Cora, Citeseer and Wiki has only one label, we employ L2-regularized logistic regression (L2R-LR), the default setting of Liblinear~\cite{fan2008liblinear} to build classifiers. For multi-label classification in BlogCatalog, we train one-vs-rest logistic regression, and employ \emph{micro-F1} for evaluation.

For link prediction, we employ a standard evaluation metric \textbf{AUC}~\cite{hanley1982meaning}. Given the similarity of all vertex pairs, AUC is the probability that a random unobserved link has higher similarity than a random nonexistent link. Assume that we draw $n$ independent comparisons, the AUC value is $(n_1+0.5n_2)/n$, where $n_1$ is the times that unobserved link has a higher score and $n_2$ is the times that they have an equal score.

For community detection, we employ \textbf{modified modularity}~\cite{zhang2015incorporating} to evaluate the quality of detected overlapping communities.

\subsection{Vertex Classification}

In Table~\ref{table:accuracy_cora}, Table~\ref{table:accuracy_citeseer}, Table~\ref{table:accuracy_wiki}, and Table~\ref{table:accuracy_blog}, we show the vertex classification results under different training ratios. For each training ratio, we randomly select vertices as training set and the rest as the test set. Note that, we employ smaller training ratios on BlogCatalog to accelerate the training speed of multi-label classifiers and evaluate the performance of CNRL under sparse scenes. From these tables, we have the following observations:

(1) The proposed CNRL model consistently and significantly outperforms all baseline methods on vertex classification. Specifically, Community-enhanced DeepWalk outperforms DeepWalk, as well as Community-enhanced node2vec outperforms node2vec. It states the importance of incorporating community information and the flexibility of CNRL to various models. Moreover, with the consideration of community structure, CNRL is able to learn more meaningful and discriminative network representations and the learnt representations are suitable for predictive tasks.

(2) While MNMF performs poorly on Wiki and BlogCatalog, CNRL performs reliably on all datasets. Furthermore, CNRL achieves more than 4 points improvements than MNMF, although they both incorporate community information into NRL. It indicates that CNRL integrates community information more efficiently and has more stable performance than MNMF.

(3) With half less training data, CNRL still outperforms the baseline methods on different datasets. It demonstrates that CNRL can better handle the sparse situation.

%

To summarise, our proposed CNRL effectively encodes the community structure into vertex representations and achieves significant improvements compared with typical NRL methods on vertex classification. Besides, CNRL is flexible to various social networks, whether they are sparse or dense. Moreover, compared with typical NRL methods, it can obtain a better performance with much less labeled training data.

\textbf{Parameter Sensitivity} To verify the influence of the parameter in CNRL on vertex classification, e.g. the number of communities $K$, we conduct experiments on three datasets (e.g., Cora, Citeseer, and Wiki.) to analyze parameter sensitivity. Here, we employ the best-performed S-n2v and set the training ratio (TR) to $0.1$, $0.5$ and $0.9$ respectively and vary the number of communities from $20$ to $120$. As shown in Fig. \ref{fig:vc_sen}, our algorithm CNRL has a stable performance on vertex classification when the number of communities varies from $20$ to $120$. It indicates that our model can adapt to detect various numbers of communities during NRL. It also demonstrates the robustness of our model.

\subsection{Link Prediction}
\begin{figure*}[!htb]
\centering
\includegraphics[width=1.2\columnwidth]{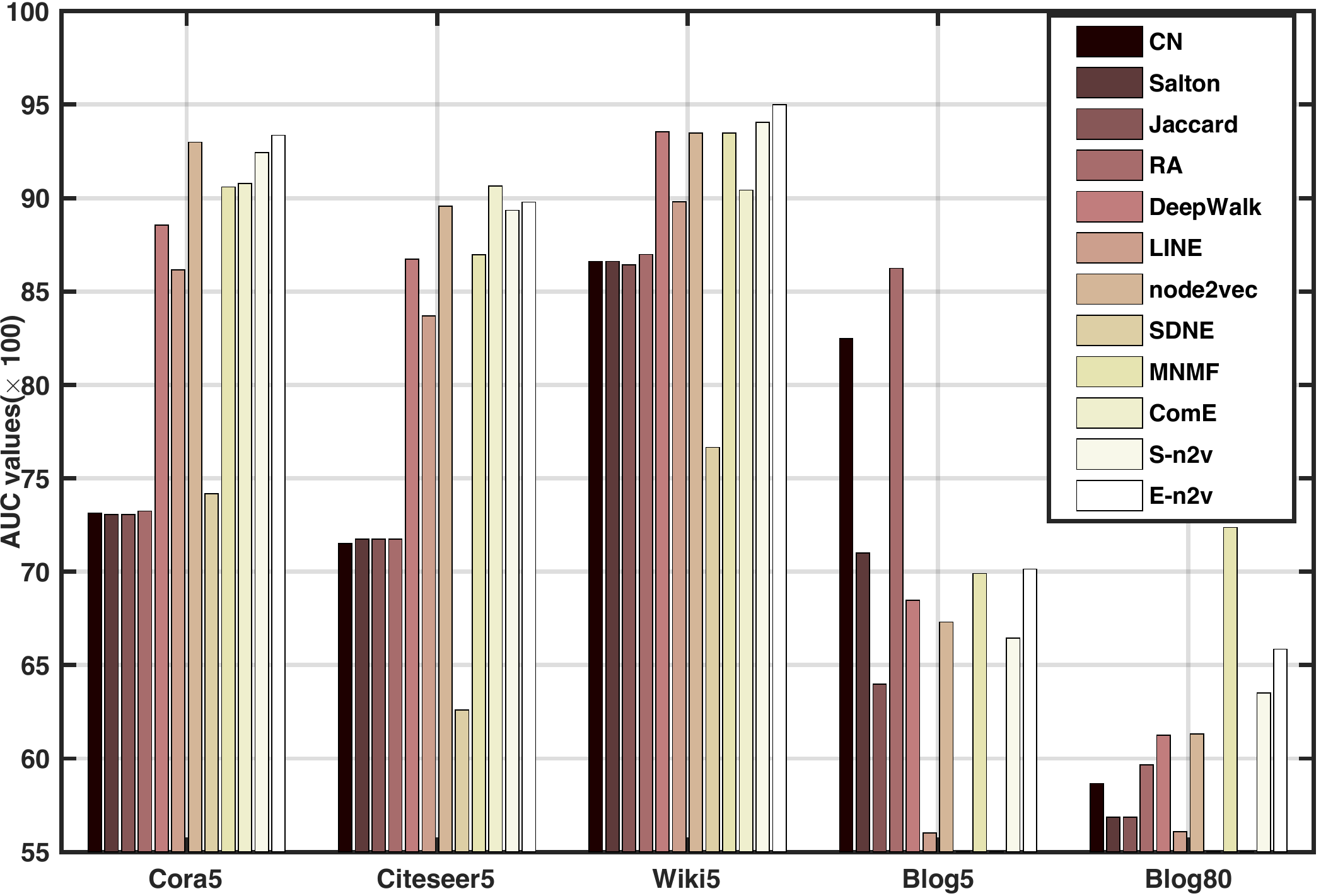}
\caption{Link prediction results when removing different amounts of links.}
\label{fig:lp_n2v}
\end{figure*}

\begin{figure*}[!htb]
\centering
\begin{minipage}{\textwidth}
\subfigure[AUC values on Cora]{
\includegraphics[width=0.31\textwidth,height=1.8in]{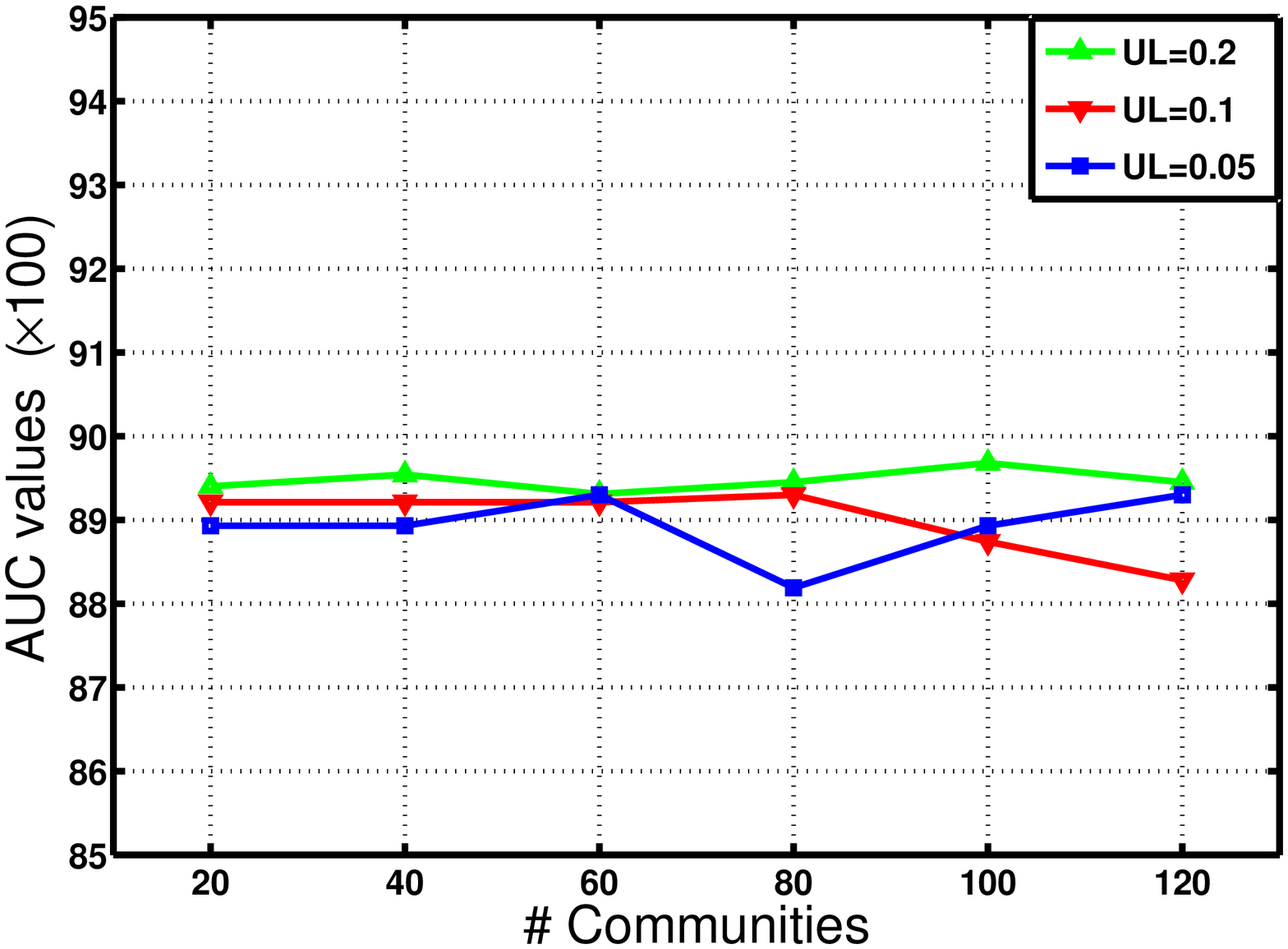}
}
\subfigure[AUC values on Citeseer]{
\includegraphics[width=0.31\textwidth,height=1.8in]{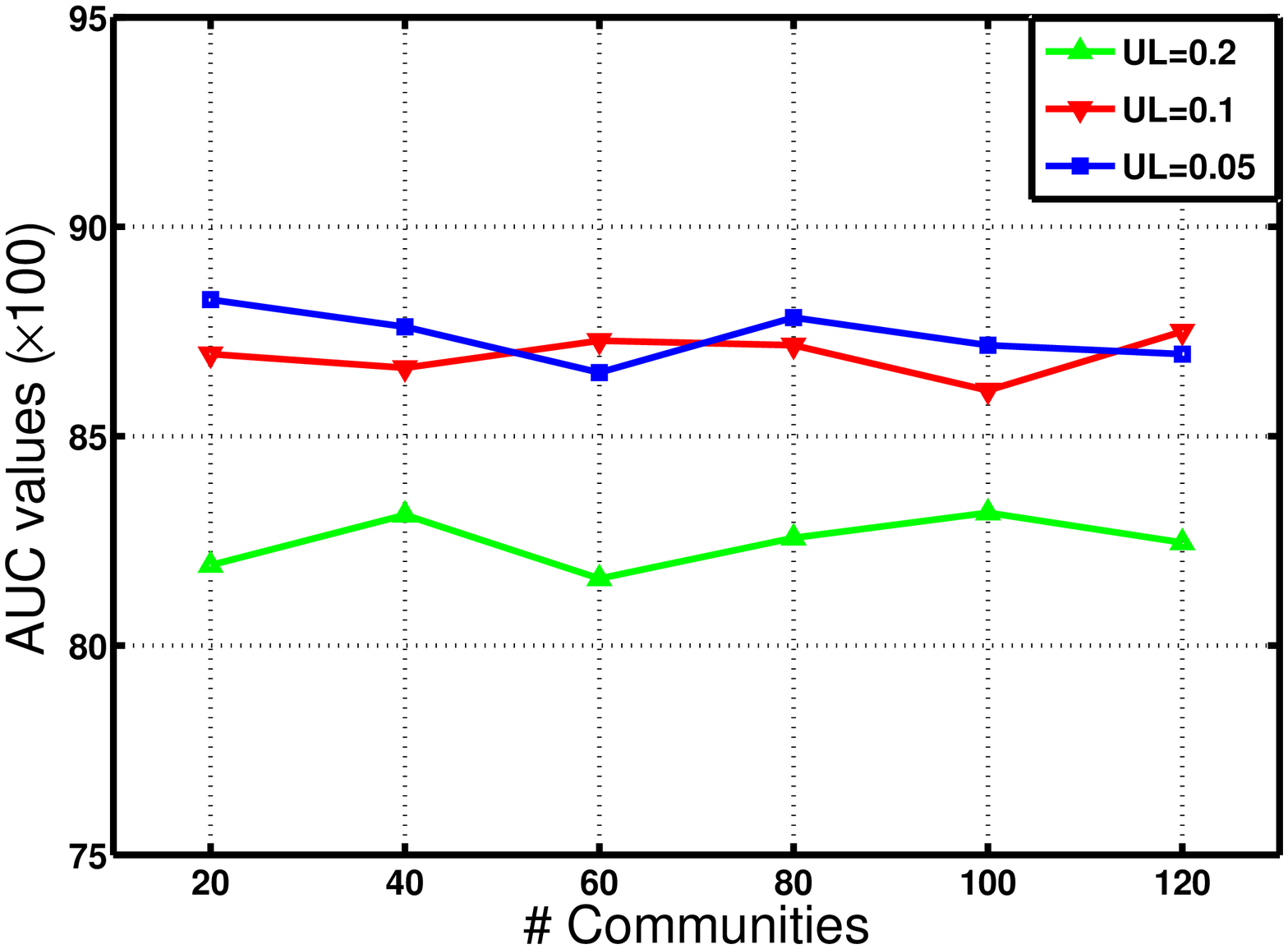}
}
\subfigure[AUC values on Wiki]{
\includegraphics[width=0.31\textwidth,height=1.8in]{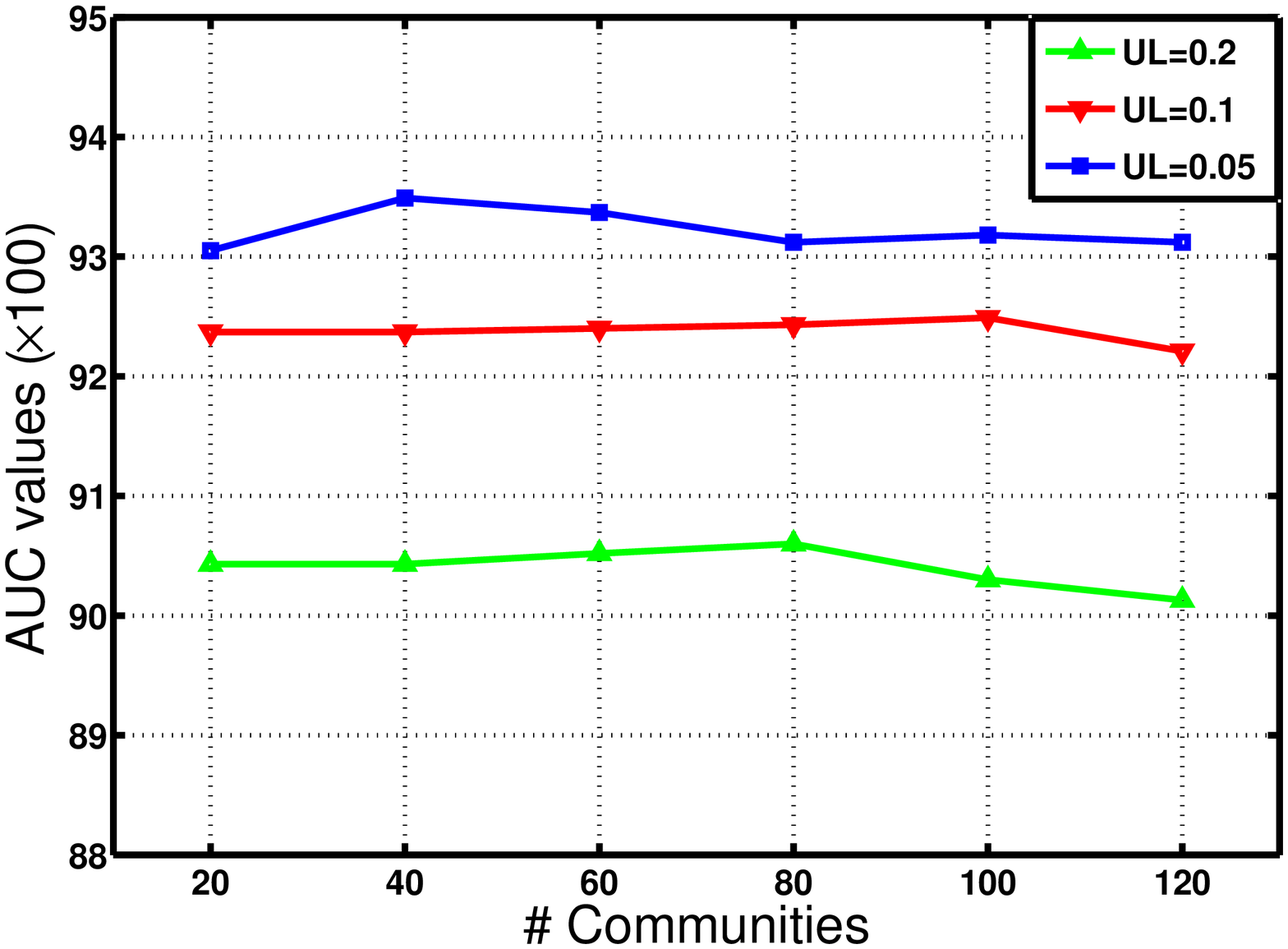}
}
\end{minipage}
\caption{Parameter sensitivity on link prediction. (UL denotes unobserved links.)}
\label{fig:lp_sen}
\end{figure*}

Community modeling models should have the ability to predict links correctly. Therefore, we employ link prediction task to evaluate our proposed Community-enhanced NRL model.

Given a network, we randomly remove a portion of links as the test set and the rest as the training set. We utilize the observed links to learn vertex representations and employ these representations to measure the similarity between two vertices, which can be further applied to predict potential links between vertices. 

We choose two kinds of baselines: link prediction methods and representation learning methods. All algorithms need to score the similarity between two vertices. Note that potentially connected vertices are expected to have higher similarity score than irrelevant ones.

We employ a standard link prediction metric, AUC\cite{hanley1982meaning}, to evaluate baselines and our method. Given the similarity of all vertex pairs, AUC is the probability that a random unobserved link has higher similarity than a random nonexistent link. Assume that we draw $n$ independent comparisons, the AUC value is
\begin{equation}
AUC=\frac{n_1+0.5n_2}{n}
\end{equation}
where $n_1$ is the times that unobserved link has a higher score and $n_2$ is the times that they have an equal score.

In Fig.~\ref{fig:lp_n2v}, we show the AUC values of link prediction on different datasets while removing $5\%$ edges. Note that, we show the results of LINE-1st on BlogCatalog, as it outperforms LINE. From this table, we observe that:

(1) In most cases, NRL methods outperform traditional hand-crafted link prediction methods. It proves that NRL methods are effective to encode network structure into real-valued representation vectors. In this case, our proposed CNRL model consistently outperforms other NRL methods on different datasets. The results demonstrate the reasonability and effectiveness of considering community structure again.

(2) For BlogCatalog, the average degree (i.e., $32.39$) of vertices is much larger than other networks, which will benefit simple statistic-based methods, such as CN and RA. Nevertheless, as shown in Fig.~\ref{fig:lp_n2v}, when the network turns sparse by removing $80\%$ edges, the performance of these simple methods will badly decrease ($25\%$ around). On the contrary, CNRL only decreases about $5\%$. It indicates that CNRL is more robust to the sparsity issue.

(3) We propose two implementations of CNRL to assign community label for vertices, denoted as statistics-based CNRL (S-CNRL) and embedding-based CNRL (E-CNRL). Regardless of the proportions of removed links, S-CNRL outperforms E-CNRL on the citation networks of Cora and Citeseer, while E-CNRL performs better than S-CNRL on the web page network Wiki. It states the necessity and rationality of the proposed two instantiations of CNRL. It makes CNRL suitable to different kinds of networks.

Observations above demonstrate the effectiveness and robustness of CNRL for incorporating community structure into vertex representations. It achieves consistent improvements than other state-of-the-art NRL methods on link prediction task. Moreover, with the two well-designed implementations, CNRL is flexible to various kinds of social networks.

\textbf{Parameter Sensitivity} We also perform experiments to analyze the influence of community size $K$ on link prediction. Same to the operations in vertex classification, we also employ S-n2v and vary community size $K$ from $20$ to $120$ and plot the AUC curves in Fig. \ref{fig:lp_sen}. As shown in this figure, our algorithm CNRL has a stable performance when the community size changes from $20$ to $120$. To be specific, the AUC values range within $2\%$ on all the three datasets, which demonstrates that the performance of our model is insensitive to the settings of the hyper-parameter $K$. It indicates the robustness and flexibility of our model.

\subsection{Community Detection}
\begin{table*}[!htb]
    \centering
    \caption{Community detection results.}
    \label{table:cd}
    \begin{tabular}{c|c|c|c|c|c|c|c|c}
        \toprule
        Datasets     & SCP & LC & MDL & BigCLAM & S-DW & E-DW & S-n2v & E-n2v\\ \midrule
        Cora & 0.076 & 0.334 & 0.427 & 0.464 & 0.464 & \textbf{1.440} &  0.447 & 1.108        \\
        Citeseer    & 0.055 & 0.315 & 0.439 & 0.403 & 0.486 & \textbf{1.861} & 0.485 & 1.515   \\
        Wiki   & 0.063  & 0.322  & 0.300 & 0.286  & 0.291 & \textbf{0.564} & 0.260 & 0.564 \\ \bottomrule
    \end{tabular}
\end{table*}
\begin{figure*}[!htb]
\centering
\begin{minipage}{\textwidth}
\subfigure{\includegraphics[width=0.33\textwidth]{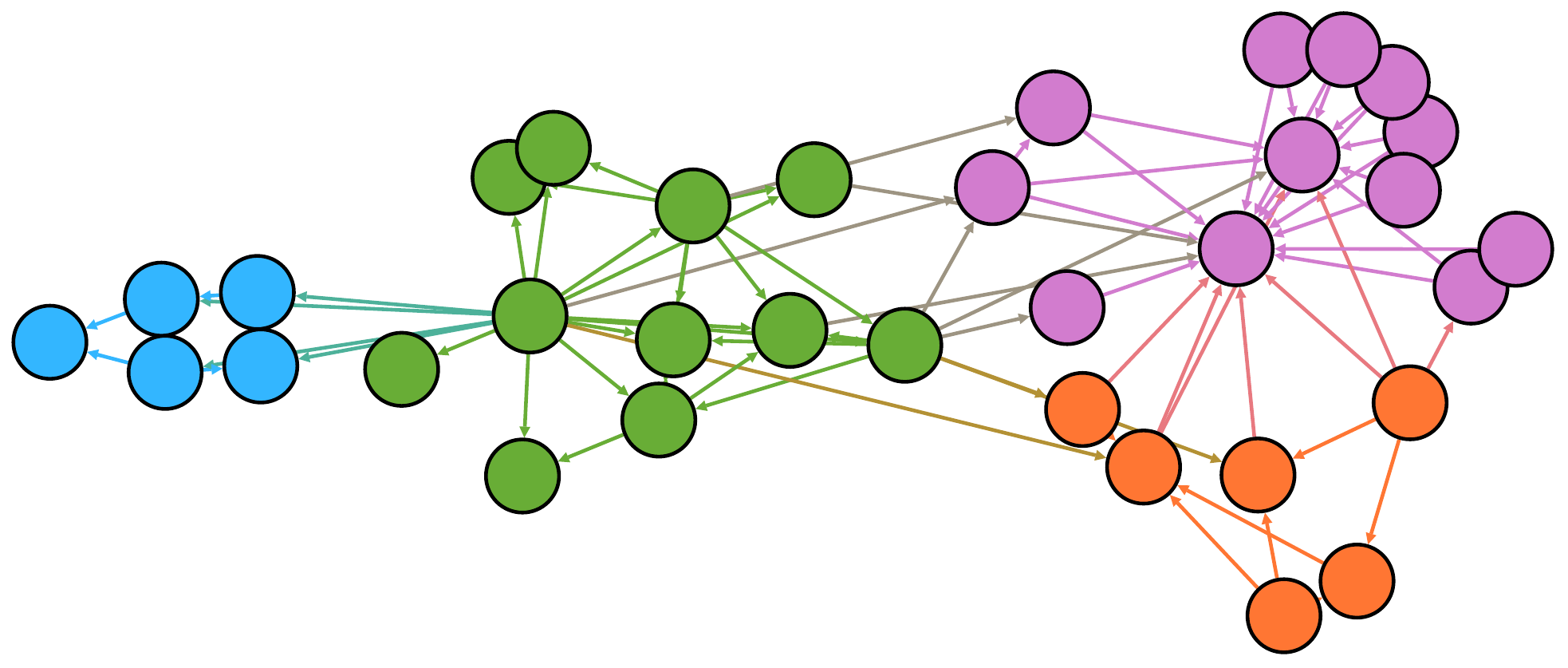}}
\subfigure{\includegraphics[width=0.33\textwidth]{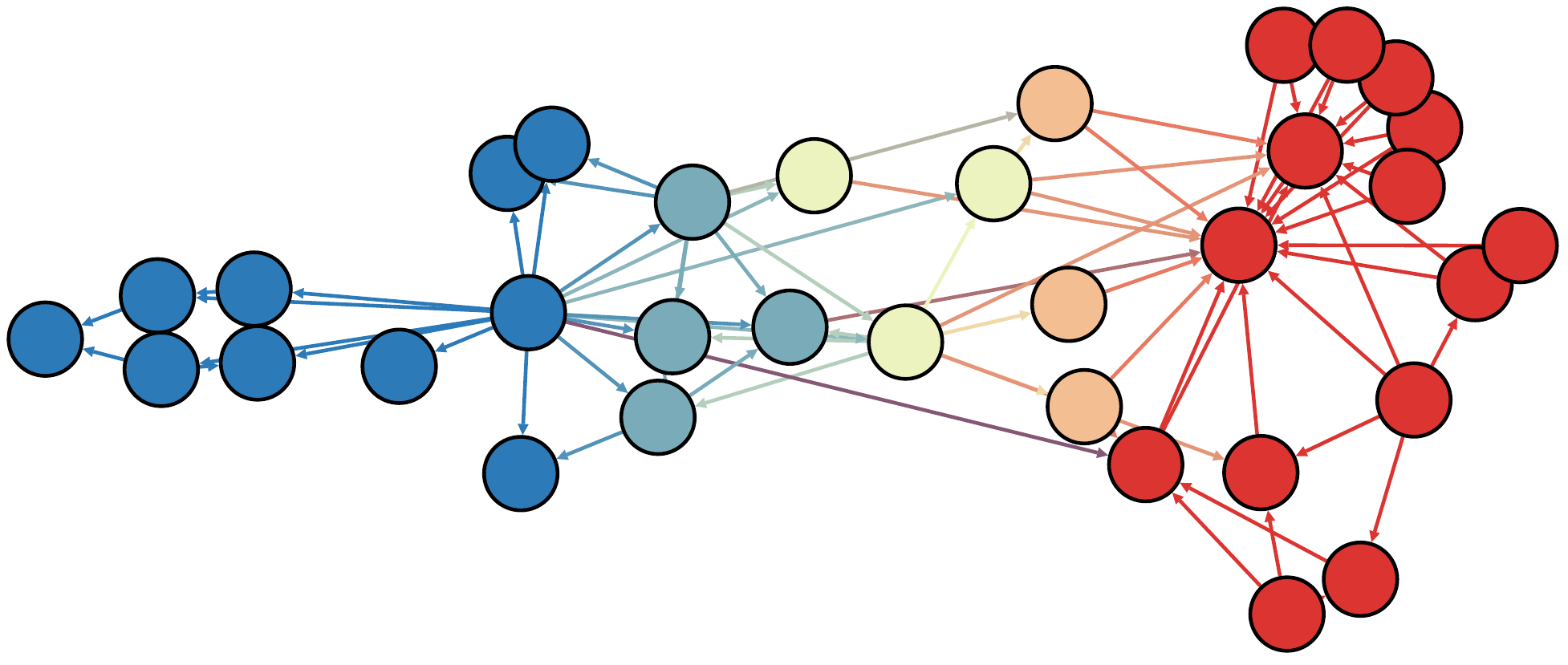}}
\subfigure{\includegraphics[width=0.33\textwidth]{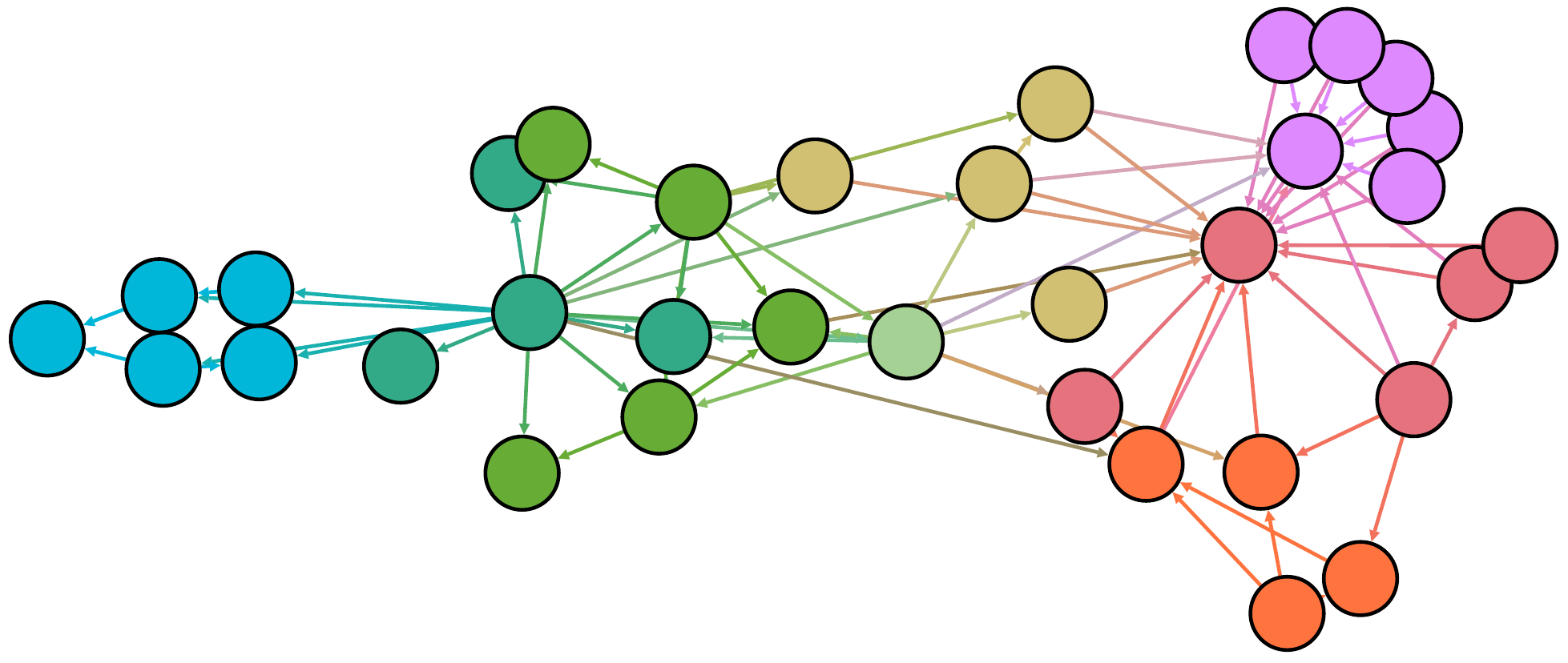}}
\end{minipage}
\caption{Detected communities on Karate (Fast Unfolding, 2 communities by CNRL, 4 communities by CNRL).}
\label{fig:karate_community}
\end{figure*}

\begin{table*}[!htb]
\centering
\caption{Top-$5$ nearest neighbors of the paper ``Protein Secondary Structure Modelling with Probabilistic Networks'' by DeepWalk and CNRL}
\label{table:case_neighbors}
\begin{tabular}{p{1.4\columnwidth}|c}
\toprule
\multicolumn{2}{c}{\textbf{CNRL}}\\ \midrule
Using Dirichlet Mixture Priors to Derive Hidden Markov Models for Protein Families & Neural Networks\\\midrule
Optimal Alignments in Linear Space using Automaton-derived Cost Functions & Neural Networks\\\midrule
Dirichlet Mixtures: A Method for Improving Detection of Weak but Significant Protein Sequence Homology & Neural Networks\\\midrule
Family-based Homology Detection via Pairwise Sequence Comparison & Neural Networks\\\midrule
The megaprior heuristic for discovering protein sequence patterns & Neural Networks\\ \midrule
\multicolumn{2}{c}{\textbf{DeepWalk}}\\ \midrule
An Optimal Weighting Criterion of Case Indexing for Both Numeric and Symbolic Attributes & Case Based\\\midrule
Using Dirichlet Mixture Priors to Derive Hidden Markov Models for Protein Families & Neural Networks\\\midrule
On The State of Evolutionary Computation & Genetic Algorithms\\\midrule
Optimal Alignments in Linear Space using Automaton-derived Cost Functions & Neural Networks\\\midrule
On Biases in Estimating Multi-Valued Attributes & Rule Learning\\ \bottomrule
\end{tabular}
\end{table*}

As there is no ground-truth communities of these datasets, we use modified modularity~\cite{zhang2015incorporating} to evaluate the quality of detected communities. Since our models and BigCLAM obtain community distribution of each vertex, we simply set a threshold $\tau = 0.1$ to select communities of each vertex (i.e., keep communities whose probability is larger than $\tau$). From Table~\ref{table:cd}, we observe that S-CNRL (S-DW or S-n2v) is comparable with other state-of-the-art community detection methods, while E-CNRL (E-DW or E-n2v) significantly outperforms these baselines. It states that the communities detected by CNRL are meaningful under the measurement of community quality. Moreover, it conforms to our assumptions about community assignment.

To summarize, all the results demonstrate the effectiveness and robustness of CNRL for incorporating community structure into vertex representations. It achieves consistent improvements comparing with other NRL methods on all network analysis tasks.

\subsection{Scalability Test}
\begin{figure}[!htb]
\centering
\includegraphics[width=0.7\columnwidth]{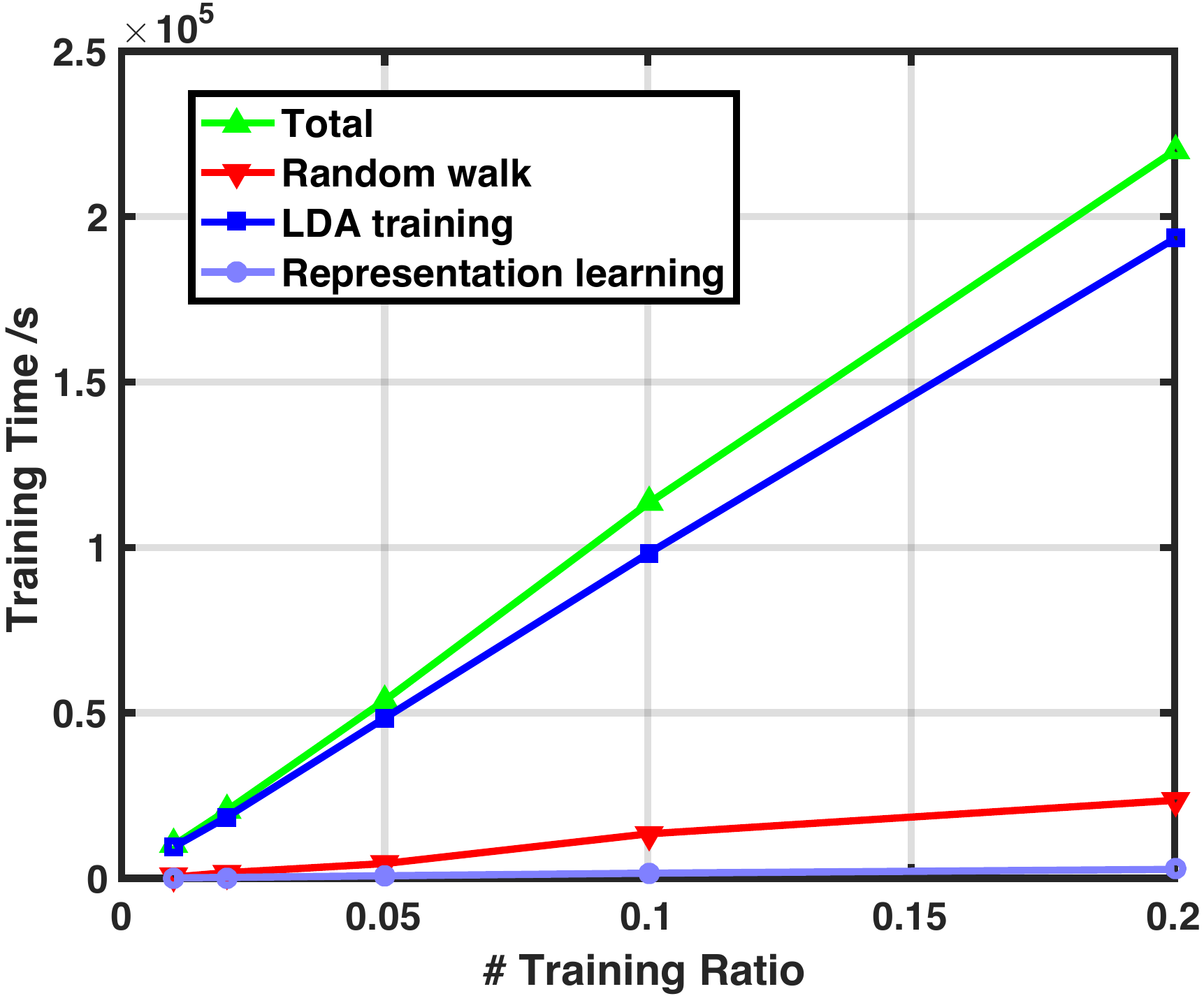}
\centering
\caption{Training time of S-n2v under various training ratios.}
\label{fig:running_time}
\end{figure}
To verify the efficiency of our model regarding the size of networks, we implement the best-performed S-n2v on a large-scale network dataset Youtube\footnote{http://socialnetworks.mpi-sws.org/data-imc2007.html}, which contains $1,138,499$ vertices and $2,990,443$ edges. Specifically, we randomly sample various proportions of vertices and test the training time of S-n2v on these sub-graphs. As shown in Fig.~\ref{fig:running_time}, S-n2v costs linear time with the number of vertices. This is mainly because most of the time cost is used for training LDA models, which is linear with the number of vertices as mentioned in the complexity analysis part. Therefore, how to accelerate the training speed of LDA models is expected to increase the efficiency of our model.
 
\subsection{Visualizations of Detected Communities}
For a more intuitive sense of detected communities, we visualize the detected overlapping communities by CNRL on a toy network named Zachary's Karate network~\cite{zachary1977information} in Fig.~\ref{fig:karate_community}. For comparison, we also show the detected non-overlapping communities by a typical community detection algorithm, Fast Unfolding~\cite{vincent2008fast}. Note that, we mark different communities with different colors, and use gradient color to represent the vertices belonging to multiple communities. From Fig.~\ref{fig:karate_community}, we observe that:

(1) CNRL is able to detect community structure with multiple scales, rather than clustering or partitioning vertices into fixed communities. Both the $2$-community version and $4$-community one are straightforward and reasonable according to the network structure.

(2) CNRL is well versed dealing with the overlapping issues in community detection. It can accurately identify vertices on community boundaries and balance the weights of the communities they belong to.

Note that, in CNRL we assume the vertex sequences reserve global community patterns, and the communities are detected with vertex sequences. Results from Fig.~\ref{fig:karate_community} conform to our intuition and verify the assumption.

\subsection{Case Study}
To demonstrate the significance of CNRL and give an intuitive experience on communities, we conduct three representative case studies as follows.
\subsubsection{Nearest neighbors}
\begin{table*}[!htb]

\centering
\caption{Community assignments and representative vertices.}
\label{table:case_communities}
\begin{tabular}{p{1.35\columnwidth}|p{0.35\columnwidth}}
\toprule
\multicolumn{2}{c}{\textbf{Community 1 (weight = 0.56)}}\\ \midrule
Learning to Act using Real-Time Dynamic Programming & Reinforcement Learning \\ \midrule
Generalized Markov Decision Processes: Dynamic-programming and Reinforcement-learning Algorithms & Reinforcement Learning \\ \midrule
On the Convergence of Stochastic Iterative Dynamic Programming Algorithms & Reinforcement Learning \\ \midrule
\multicolumn{2}{c}{\textbf{Community 2 (weight = 0.20)}}\\ \midrule
The Structure-Mapping Engine: Algorithm and Examples & Case Based \\ \midrule
Case-based reasoning: Foundational issues, methodological variations, and system approaches &  Case Based\\ \midrule
Concept Learning and Heuristic Classification in Weak-Theory Domains & Case Based \\ \midrule
\multicolumn{2}{c}{\textbf{Community 3 (weight = 0.12)}}\\ \midrule
Learning to Predict by the Methods of Temporal Differences & Reinforcement Learning \\\midrule
Generalization in Reinforcement Learning: Safely Approximating the Value Function & Reinforcement Learning \\ \midrule
Exploration and Model Building in Mobile Robot Domains & Reinforcement Learning \\ \bottomrule
\end{tabular}
\end{table*}

We provide a case in Cora to state the importance of considering community structure in NRL.
The ``paper" is titled as ``Protein Secondary Structure Modelling with Probabilistic Networks" and belongs to the research field of ``Neural Networks''.
With learnt representation vectors by DeepWalk and S-CNRL, we measure the distance between vertices with cosine similarity and give the top-$5$ nearest neighbors in Table \ref{table:case_neighbors}.

From Table \ref{table:case_neighbors}, the most straightforward observation is that CNRL gives $5$ nearest neighbors with the same labels, while DeepWalk only gives $2$.
It's reasonable as vertices in a community attempt to share the same attributes. The attributes are always hidden in the communities, which are utilized by CNRL.
More detailed observations can be achieved by analyzing their topics. The provided ``paper" is related to topics including ``protein",``structure" and ``probabilistic networks".
Most of the neighbors found by DeepWalk are unrelated to these topics. In contrast, most of the neighbors found by CNRL are closely correlated with them.
More specifically, the nearest neighbors found by CNRL can be roughly assigned into ``protein" related and ``probabilistic network" related categories,
which indicates the existence of latent correlations with community structure.

\subsubsection{Community Assignments}
Compared with DeepWalk, CNRL can learn not only the community-enhanced vertex representations but also the community assignments. We provide a case in Cora and its community assignments in Table~\ref{table:case_communities}. The selected ``paper" is titled as ``Using a Case Base of Surfaces to Speed-Up Reinforcement Learning" and belongs to the field of ``Reinforcement Learning''. For each community, we follow Eq. (\ref{eq:vc_count}) to select representative vertices.

From this table, we observe that each community has its own characteristics. For example, community $1$ is related to ``Dynamic Programming", which is a sub-field in ``Reinforcement Learning".
Community $2$ is relevant to ``Cased Based" research, and community $3$ concerns more about the learning and modeling methods in ``Reinforcement Learning".
According to the title of the selected vertex, we find that it is actually relevant to all these communities and the weights can reflect its relevance to the communities.

\subsubsection{Global Community Patterns}
\begin{table}[!htb]
    \centering
    \caption{Representative words of communities in Cora}
    \label{table:communities}
    \begin{tabular}{c|l}
        \toprule
        Community ID     & Words and Frequencies  \\ \midrule
         0 & Models:13 Hidden:11 Markov:10 \\\midrule
         6 & Reasoning:13 Case-Based:13 Knowledge:7  \\\midrule
         8 & Genetic:23 Programming:16 Evolution:9 \\\midrule
         13 & Boosting:6 Bayesian:6 Classifiers:6 \\\midrule
         15 & Neural:14 Networks:11 Constructive:7 \\\midrule
         19 & Reinforcement:21 Markov:8 Decision:8 \\
        \bottomrule
    \end{tabular}
\end{table}

In this part, we attempt to make clear the insights of community patterns from an overall perspective.

We train S-CNRL on Cora when $K = 20$, and select the most related vertices according to whether $\Pr(v|c) > \eta$. Here, the threshold $\eta$ is set to $0.005$. Afterwards, we identify the most-frequent words in titles for each community. We ignore ``Machine" and ``Learning" as Cora is a machine learning paper set and the titles of most papers contain these words. Due to space limitations, we only list 6 typical communities in Table~\ref{table:communities}.

From this table, we get a primary experience of the division in machine learning filed. These communities are discriminative and corresponding to ``Hidden Markov Models", ``Case-based Reasoning", ``Genetic Programming", ``Neural Networks" and ``Reinforcement Learning" respectively. It demonstrates that CNRL is effective in detecting high-quality communities.

\section{Conclusion and Future Work}
In this paper, we propose a novel community enhancement mechanism for NRL. By taking full consideration of the critical community information, the proposed method jointly embeds local vertex characteristics and global community patterns into vertex representations, as well as simultaneously detecting community structure. Moreover, CNRL can be easily implemented on conventional random walk based NRL models, e.g. DeepWalk and node2vec. It achieves significant improvements compared with existing state-of-the-art NRL algorithms when applying the learnt representations to network analysis tasks. Besides, CNRL can effectively detect overlapping communities on multiple scales.

In future, we will explore the following directions:

(1) In this work, we employ topic modeling methods on random walk sequences for community detection inspired by the analogy between natural language and networks. However, the correlation between topics and communities still lacks theoretical demonstration. Therefore, we aim to prove this analogy mathematically through matrix factorization.

(2) We seek to investigate the extensibility of our model on incorporating heterogeneous information in social networks, such as text contents and attributes. Our method is expected to be flexible to incorporate the heterogeneous information comprehensively. 

(3) Another intriguing direction would be semi-supervised learning of our model. We will adapt the representation learning for specific tasks such as vertex classification and take label information of training set into account to enhance the quality of vertex representations for prediction.

\section*{Acknowledgements}
This work is supported by the National Natural Science Foundation of China (NSFC No. 61772302, 61532010, 61661146007). This research is also part of the NExT++ project, supported by the National Research Foundation, Prime Minister’s Office, Singapore under its IRC@Singapore Funding Initiative. Tu is also supported by China Postdoctoral Innovative Talent Support Programme.
\bibliographystyle{IEEEtran}

\bibliography{community}
\vspace{-5em}
\begin{IEEEbiography}[{\includegraphics[width=1in,height=1.25in,clip,keepaspectratio]{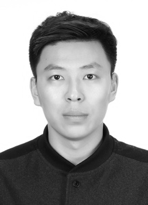}}]{Cunchao Tu} is a postdoc of the Department of Computer Science and Technology, Tsinghua University. He got his BEng degree in 2013 from the Department of Computer Science and Technology, Tsinghua University. His research interests are natural language processing and social computing. He has published several papers in international conferences and journals including ACL, IJCAI, EMNLP, COLING, and ACM TIST.
\end{IEEEbiography}
\vspace{-5em}
\begin{IEEEbiography}[{\includegraphics[width=1in,height=1.25in,clip,keepaspectratio]{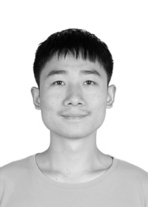}}]{Xiangkai Zeng} is an undergraduate student in the School of Computer Science and Engineering, Beihang University, Beijing, China. His research interests are natural language processing and deep learning.
\end{IEEEbiography}
\vspace{-5em}
\begin{IEEEbiography}[{\includegraphics[width=1in,height=1.25in,clip,keepaspectratio]{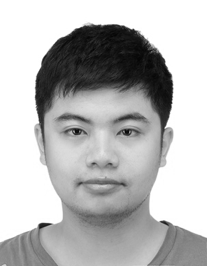}}]{Hao Wang} is a PhD student in the School of Physical and Mathematical Sciences, Nanyang Technological University. He got his BEng degree in 2017 from the Department of Computer Science and Technology, Tsinghua University. His research interests include algorithmic game theory, mechanism design, and social network analysis.\end{IEEEbiography}
\vspace{-4em}
\begin{IEEEbiography}[{\includegraphics[width=1in,height=1.25in,clip,keepaspectratio]{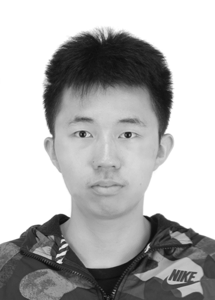}}]{Zhengyan Zhang} is an undergraduate student of the Department of Computer Science and Technology, Tsinghua University. His research interests include natural language processing and social computing.\end{IEEEbiography}
\vspace{-5em}
\begin{IEEEbiography}[{\includegraphics[width=1in,height=1.25in,clip,keepaspectratio]{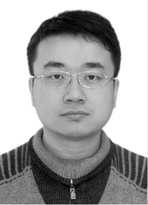}}]{Zhiyuan Liu} is an associate professor of the Department of Computer Science and Technology, Tsinghua University. He got his BEng degree in 2006 and his Ph.D. in 2011 from the Department of Computer Science and Technology, Tsinghua University. His research interests are natural language processing and social computation. He has published over 40 papers in international journals and conferences including ACM Transactions, IJCAI, AAAI,  ACL and EMNLP. 
\end{IEEEbiography}
\vspace{-4em}
\begin{IEEEbiography}[{\includegraphics[width=1in,height=1.25in,clip,keepaspectratio]{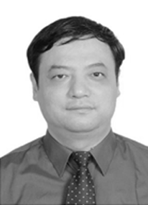}}]{Maosong Sun} is a professor of the Department of Computer Science and Technology, Tsinghua University. He got his BEng degree in 1986 and MEng degree in 1988 from Department of Computer Science and Technology, Tsinghua University, and got his Ph.D. degree in 2004 from Department of Chinese, Translation, and Linguistics, City University of Hong Kong. His research interests include natural language processing, Chinese computing, Web intelligence, and computational social sciences. He has published over 150 papers in academic journals and international conferences in the above fields. He serves as a vice president of the Chinese Information Processing Society, the council member of China Computer Federation, the director of Massive Online Education Research Center of Tsinghua University, and the Editor-in-Chief of the Journal of Chinese Information Processing.
\end{IEEEbiography}
\vspace{-3em}
\begin{IEEEbiography}[{\includegraphics[width=1in,height=1.25in,clip,keepaspectratio]{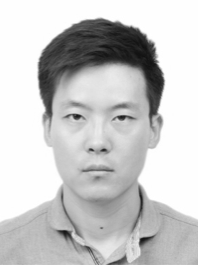}}]{Bo Zhang} is a researcher of the Search Product Center, WeChat Search Application Department, Tencent Inc., China. He got his BEng degree in 2009 from School of Computer Science, Xidian University, China, and got his Master Degree in 2012 from School of Computer Science and Technology, Zhejiang University, China.
\end{IEEEbiography}
\vspace{-3.5em}
\begin{IEEEbiography}[{\includegraphics[width=1in,height=1.25in,clip,keepaspectratio]{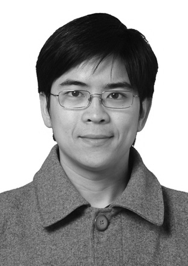}}]{Leyu Lin} is a researcher of the Search Product Center, WeChat Search Application Department, Tencent Inc., China. He got his BEng degree in 2004 from Department of Computer Science and Technology, Tianjin Normal University, China.
\end{IEEEbiography}
\end{document}